\documentclass[aps,prx,reprint,groupedaddress]{revtex4-2}
\usepackage{comment}
\usepackage{graphicx}% Include figure files
\usepackage{dcolumn}% Align table columns on decimal point
\usepackage{bm}% bold math
\usepackage{xcolor}
\usepackage{float}
\usepackage{amssymb,amsmath,latexsym,upgreek,mathtools}
\newcommand\numberthis{\addtocounter{equation}{1}\tag{\theequation}}

\usepackage{xr}
\newcommand{\Gn}{G_N}
\newcommand{\gn}{g_N}
\newcommand{\rhon}{\rho_{N}}

\newcommand{\rhot}{\rho_{\text{T}}}
\newcommand{\rhoRNA}{\rho_{\text{RNA,T}}}

\newcommand{\rhoRNAc}{\rho_{\text{RNA}}^{\text{c}}}
\newcommand{\rhoRNAbg}{\rho_{\text{RNA}}^{\text{bg}}}

\newcommand{\rhorf}{\rho_{\text{RNA,f}}}

\newcommand{\rhoonea}{\rho_1^{\alpha}}

\newcommand{\rhoonebg}{\rho_1^{\text{bg}}}
\newcommand{\rhoonec}{\rho_1^{\text{c}}}

\newcommand{\rhonma}[4]{\rho_{\text{#1},(#2,#3)}^{#4}}

\newcommand{\kt}{k_{\text{B}}T}
\newcommand{\gads}{g_{\text{ads}}}

\newcommand{\Vc}{V_{\text{c}}}
\newcommand{\Rc}{R_{\text{c}}}
\newcommand{\Vbg}{V_{\text{bg}}}
\newcommand{\Vr}{V_{\text{r}}}
\newcommand*{\VT}{V_\text{tot}}

\newcommand{\Kc}{K_{\text{c}}}

\newcommand{\KRNA}{K_{\text{RNA}}}
\newcommand{\Kcog}{K_{\text{cog}}}
\newcommand{\KNC}{K_{\text{NC}}}

\newcommand{\fcog}{f_{\text{cog}}}

\newcommand{\rhonut}{\rho_{\nu,\text{T}}}

\newcommand{\rhonct}{\rho_{\text{NC},\text{T}}}

\newcommand{\rhonnua}{\rho_{N}^{\nu,\alpha}}
\newcommand{\rhonnubg}{\rho_{N}^{\nu,\text{bg}}}
\newcommand{\rhonnuc}{\rho_{N}^{\nu,\text{c}}}

\newcommand{\rhoncog}{\rho_{N}^{\text{cog}}}
\newcommand{\rhocogt}{\rho_{\text{cog},\text{T}}}

\newcommand{\rhonnc}{\rho_{N}^{\text{NC}}}

\newcommand{\cSS}{\rho_\text{SS}}

\newcommand{\yield}{x_{\text{enc}}^{\text{T}}}
\newcommand{\yieldcog}{x_{\text{enc}}^{\text{cog}}}
\newcommand{\yieldnc}{x_{\text{enc}}^{\text{NC}}}

\newcommand{\yieldBGnu}{x_{\text{enc}}^{\nu,\text{bg}}}
\newcommand{\yieldCnu}{x_{\text{enc}}^{\nu,\text{c}}}
\newcommand{\cac}{\rho_{\text{CAC}}}
\newcommand{\selec}{s}
\newcommand{\selecnorm}{\bar{s}}

\newcommand{\fElong}{f_0}

\newcommand{\kads}{k_{\text{ads}}}
\newcommand{\kdes}{k_{\text{desorb}}}

\newcommand{\nnuc}{n_{\text{nuc}}}

\newcommand{\ktrans}{k_{\text{tr}}}

\newcommand{\adsratesub}{\mathcal{N}_1^{\alpha}}

\newcommand{\diffratesub}{\mathcal{D}_1^{\alpha}}
\newcommand{\diffratesubc}{\mathcal{D}_1^{\text{c}}}

\newcommand{\transratesub}{\mathcal{T}^{\alpha}}

\newcommand{\adsraterna}{\mathcal{N}_{\nu,(n,m)}^{\alpha}}
\newcommand{\assemraterna}{\mathcal{M}_{\nu,(n,m)}^{\alpha}}
\newcommand{\diffraterna}{\mathcal{D}_{\nu,(n,m)}^{\alpha}}
\newcommand{\diffraternaopt}[2]{\mathcal{D}_{#1,(n,m)}^{#2}}

\newcommand{\kdlr}{k_{\text{DL, RNA}}}
\newcommand{\kdls}{k_{\text{DL, 1}}}

\newcommand{\gNuc}{g_\text{nuc}}
\newcommand{\gElong}{g_\text{elong}}

\newcommand{\dee}{\text{d}}

\begin{document}

\title{Liquid-liquid phase separation enables highly selective viral genome packaging}

\author{Layne B. Frechette}
\author{Michael F. Hagan}
\email[]{hagan@brandeis.edu}

\affiliation{Martin Fisher School of Physics, Brandeis University, Waltham, Massachusetts 02453, USA}

\date{\today}

\begin{abstract}
In many viruses, hundreds of proteins assemble an outer shell (capsid) around the viral nucleic acid to form an infectious virion. How the assembly process selects the viral genome amidst a vast excess of diverse cellular nucleic acids is poorly understood. It has recently been discovered that many viruses perform assembly and genome packaging within liquid-liquid phase separated biomolecular condensates inside the host cell. However, the role of condensates in genome packaging is poorly understood. Here, we construct equilibrium and dynamical rate equation models for condensate-coupled assembly and genome packaging. We show that when the viral genome and capsid proteins favorably partition into the condensate, assembly rates, yields, and packaging efficiencies can increase by orders of magnitude. Selectivity is further enhanced by the condensate when capsid proteins are translated during assembly and packaging. Our results suggest that viral condensates provide a mechanism to ensure robust and highly selective assembly of virions around viral genomes. More broadly, our results may apply to other types of selective co-assembly processes that occur within biomolecular condensates, and suggest that liquid-liquid phase-separated condensates could be exploited for selective encapsulation of microscopic cargo in human-engineered systems.
\end{abstract}

\maketitle

\section{Introduction}
Capsid assembly around the viral genome is an essential step in the lifecycle of many viruses. To form an infectious virion, the capsid proteins must selectively encapsulate the viral RNA amidst excess cellular RNA, and many do so with high specificity (e.g. 99\% \cite{Routh2012, Comas-Garcia2012}). How viruses achieve such selective packaging remains a critical unsolved problem in virology \cite{Comas-Garcia2019,Rastandeh2025}. More broadly, assembly around specific cargoes has also become a key design challenge for biotechnology, e.g. to form gene or drug delivery vectors or nanoreactors \cite{Wilkerson2018,Galaway2013,Guenther2014}. It has recently become clear that many viruses perform assembly and genome packaging within liquid-liquid phase separated biomolecular condensates \cite{Brocca2020,Etibor2021,FernandezdeCastro2021,Gaete-Argel2019,Lopez2021,Schoelz2017,Alenquer2019}. In this article, we use theoretical modeling to show that coupling of assembly and genome packaging with condensates can lead to extremely selective packaging of the viral genome. These insights shed new light on viral lifecycles and could help to identify targets for antiviral agents that interfere with viral condensates. Our models also show how to apply principles from virus assembly to these human-engineered applications, by using liquid-liquid phase separation (LLPS) to drive specific cargo packaging.

Several mechanisms have been identified for selective genome packaging. Intense experimental and theoretical research has shown that for many viruses, the genome contains one or more sequence-specific `packaging signals' that interact with `packaging sites' on capsid proteins \cite{Rao2006,Stockley2013a,Pappalardo1998,Lu2011,DSouza2005,Sorger1986,Beckett1988,Bunka2011,Dykeman2011,Comas-Garcia2019}. Furthermore, the charge and structure of viral RNA molecules is optimal for encapsidation by their capsid proteins \cite{Perlmutter2013,Erdemci-Tandogan2014, Comas-Garcia2012, Bancroft1969,Hohn1969}. Other mechanisms include coordinated capsid protein translation and assembly \cite{Dykeman2014, Annamalai2008,Kao2011}, as well as subcellular localization of viral components \cite{Bamunusinghe2011,Rao2014a,Tarquini2018,Koziel2017}. Despite this evidence, how selective packaging depends on these or other as yet unidentified mechanisms remains poorly understood for most viruses. In this article, we investigate how coupling assembly and genome packaging to condensates can significantly promote the latter two mechanisms. 

Biomolecular condensates act as `membraneless organelles' that locally concentrate specific components to facilitate cellular processes, such as signaling \cite{Jaqaman2021,Jeon2025}, the cellular stress response \cite{Grousl2022,Malinovska2012,Molliex2015,Patel2015a}, transcription \cite{Hnisz2017,Cho2018,Chong2018,Henninger2021,Sabari2018}, and cell division. Condensates can also control self-assembly within cells. Examples include the assembly of clathrin cages during endocytosis \cite{Day2021}; formation of post-synaptic densities \cite{Zeng2018a} and pre-synaptic vesicle release sites (active zones) \cite{McDonald2020, Emperador-Melero2021} at neuronal synapses; assembly of actin within polypeptide coacervates \cite{McCall2018}, and modulating aggregation of $\alpha$-synuclein \cite{Lipinski2022}.
Beyond normal cellular functions, many pathogens construct or exploit condensates within their host cells. For example, many viruses perform assembly and packaging of the viral nucleic acid within condensates known as virus factories, replication sites, Negri bodies, inclusion bodies, or viroplasms \cite{Borodavka2018,Borodavka2017,Brocca2020,Carlson2020,Etibor2021,FernandezdeCastro2021,Gaete-Argel2019,GarcesSuarez2019,Geiger2021,Guseva2020,Kieser2020,Lopez2021,Luque2020,Nikolic2017,Pan2021,Papa2021,Risso-Ballester2021,Savastano2020,Schoelz2017,Trask2012,Wang2021,Cubuk2021,He2024,Vallbracht2025}.

Condensate formation is driven by weak, multivalent interactions among the constituents leading to LLPS, although condensates are also regulated by nonequilibrium processes including internal chemical reactions \cite{Banani2017,Carenza2020,PlysAaron2018,Weber2019,Zwicker2014,Alberti2019,Bergeron-Sandoval2016,Bergeron-Sandoval2021,Brangwynne2011,Brangwynne2013,Brangwynne2015,Dignon2020,Falahati2019,Feric2016,Hyman2012a,Jacobs2021,Jacobs2017,Joseph2021,Perry2019,Sanchez-Burgos2021,Style2018,Zhou2018a,Bergeron-Sandoval2016, Hnisz2017, Molliex2015, Wu2020, Henninger2021, Sabari2018, Chong2018, Cho2018,Zwicker2014a, Zwicker2015, Zwicker2017, Zwicker2025, Weber2019c, Ziethen2023, Ziethen2024, Soding2020, Kirschbaum2021, Luo2025, Rossetto2025, Choi2019, Brangwynne2015, Ruff2019, Holla2024, Dignon2018c, Dignon2019a, Dignon2019b, Shea2021, Rekhi2024, Kapoor2024, Hnisz2017, Sabari2018, Schede2023, Banani2024,Pyo2023,GrandPre2023,Zhang2024,Grigorev2025,Jacobs2021,Chen2024,Li2024,Lin2018,Cinar2019,Pal2021,Shen2023,Das2018,Nguemaha2018,Kota2022,Zhou2024}. 
Components that drive phase separation are called scaffolds, while components that preferentially partition into condensates but do not provide the primary phase separation driving force are called clients~\cite{Ditlev2018}.

\begin{figure*}%[htbp]
    %\centering
    \includegraphics[width=\linewidth]{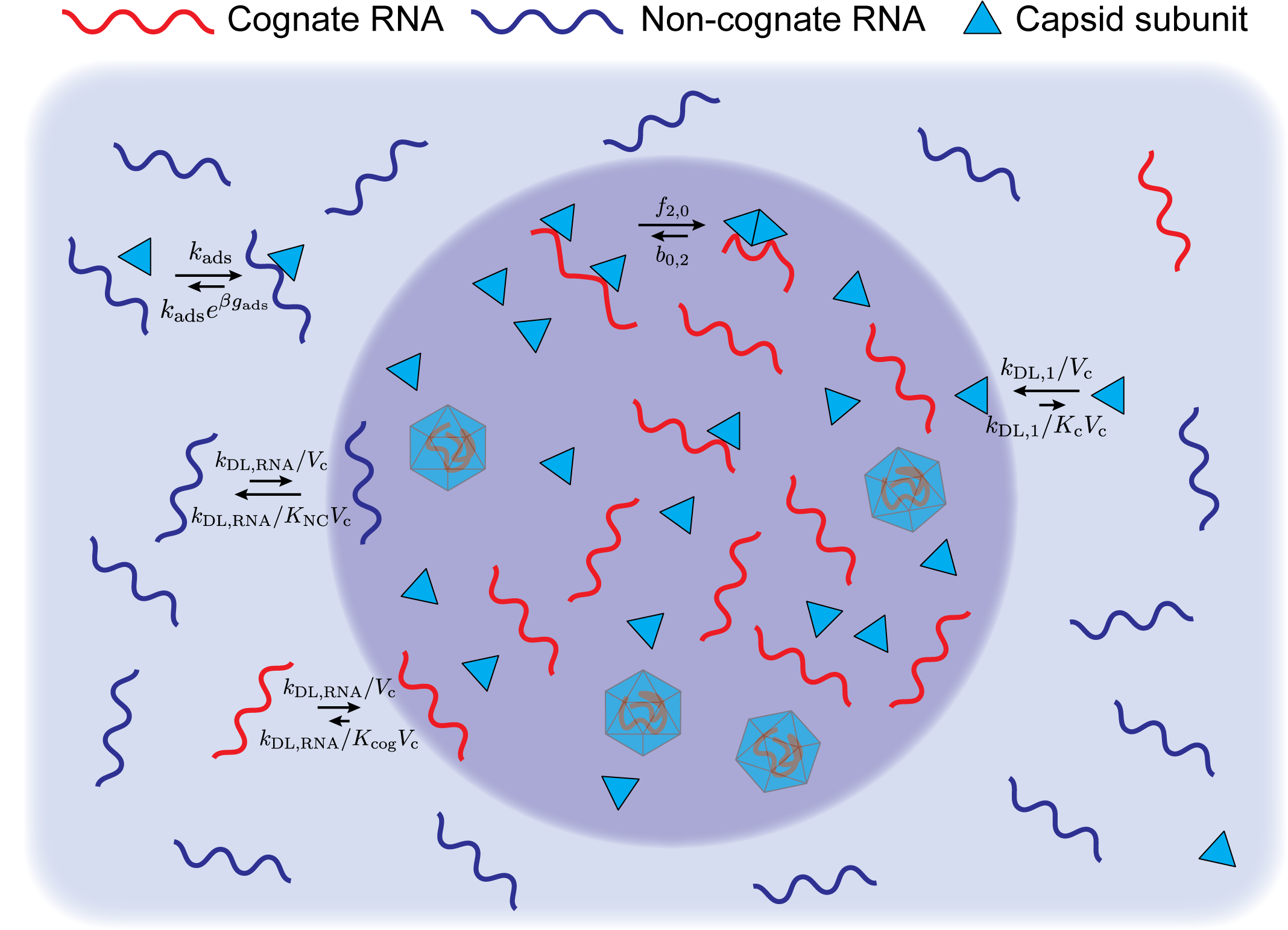}
    \caption{\textbf{Schematic of viral genome packaging within biomolecular condensates formed by liquid-liquid phase separation (LLPS).} The viral condensate (light purple circle) is enriched in viral components by recruiting viral RNA (red) and capsid subunits (cyan) while disfavoring host RNA (blue). The condensate has volume $\Vc$, while the surrounding ``background'' has volume $\Vbg=V-\Vc$. We define the volume fraction $\Vr=\Vc/\Vbg$; unless otherwise noted, $\Vr=10^{-3}$. Subunits and RNA strands enter the condensate with diffusion-limited rate constants $\kdls/\Vc$ and $\kdlr/\Vc$, respectively (Eqs.~\ref{eq:subunit_diff} and ~\ref{eq:rna_diff}), and exit with rate constants $\kdls/\Kc\Vc$ and $\kdlr/K_{\nu}\Vc$, for $\nu=\text{cog, NC}$. Subunits adsorb onto RNA with rate constant $\kads$ and binding free energy $\gads$, such that the unbinding rate constant is $\kads e^{\beta \gads}$. Adsorbed subunits can subsequently assemble; an RNA molecule with $n$ adsorbed subunits and $m$ assembled subunits undergoes dimerization (if $m=0$) or association of a subunit to the intermediate (if $m\geq 2)$ with rate $f_{n,m}$. Disassembly occurs with rate $b_{n,m}$.}   \label{fig:schematic}
\end{figure*}

In most viral condensates, favorable interactions drive preferential partitioning (i.e. increase the local concentration) of viral RNA, capsid proteins, and other components. In many negative-strand ssRNA viruses~\cite{Guseva2020,Nikolic2017,Heinrich2018,Vallbracht2025,Galloux2020}, coronaviruses~\cite{Iserman2020,Savastano2020,Milles2018}, and retroviruses~\cite{Monette2020,Monette2022}, the nucleocapsid protein is the scaffold and primary constituent of the condensate.  However, in other viruses, such as norovirus \cite{Kaundal2024}, adenovirus \cite{Hidalgo2021}, rotavirus \cite{Papa2019,Papa2021,Geiger2021} and reovirus \cite{He2024}, auxiliary viral proteins are the scaffolds.

While there have been comparatively few investigations of condensate-coupled assembly, \textit{in vitro} experiments have demonstrated that viral proteins and/or nucleic acids form condensates (e.g. \cite{BessaLuiza2022,Chen2020,Iserman2020, Guseva2020, Savastano2020}), and that condensates can accelerate assembly of capsids \cite{Guseva2020}.
On the theoretical side,
recent works using chemical kinetics-based rate equations \cite{Weber2019a, Lipinski2022, Schmit2022, Ponisch2023,Michaels2022,Weber2019,Bartolucci2024, Hagan2023} and molecular dynamics simulations \cite{Frechette2025} suggest that preferential partitioning of subunits into condensates can significantly enhance assembly rates and robustness against parameter variations. However, models have yet to include genome packaging or other forms of co-assembly within a condensate.

Here, we advance beyond these previous studies by considering the assembly of capsids around nucleic acids within a condensate. We find that the condensate can significantly increase the assembly rates and yields of capsids filled with nucleic acids. Moreover, preferential partitioning (i.e. increased local concentration) of the viral RNA into the condensate can lead to highly selective packaging, and this effect is amplified if nonviral RNA is preferentially excluded from the condensate. Under these conditions we observe essentially 100\% selectivity across a broad range of physiologically relevant conditions. Furthermore, motivated by the observation that some viruses include ribosomes in their condensates \cite{Desmet2014,Tenorio2019,FernandezdeCastro2014,Wang2022,Liu2024,Jobe2024}, we find that capsid protein translation during assembly can further increase yields and selective packaging when the condensate is present. Based on these observations, we suggest that viral condensates provide a potent mechanism for selective genome packaging. This mechanism could work in synergy with previously proposed mechanisms for selective packaging \cite{Comas-Garcia2019,Stockley2013}, such as packaging signals. More broadly, our model is sufficiently general that it could be used to design approaches that utilize LLPS to encapsulate specific cargoes in a variety of biological and nanomaterials systems.

\section{Model}
We consider a minimal model for the effect of condensates on selective packaging of RNA. We focus on ssRNA viruses, and we model the case in which nonstructural proteins are scaffolds (e.g. the norovirus protein NS7, an RNA-dependent RNA polymerase~\cite{Kaundal2024}), while the RNA and assembly subunits are clients. We assume that subunits are sufficiently dilute that their effects on condensate stability can be neglected. 
However, our model can be readily extended to the case where the capsid protein is the scaffold (which we are considering in ongoing work), and to viruses with dsRNA (e.g. rotavirus) or DNA genomes.  

We extend rate equation models that we previously developed to describe assembly of empty capsids (containing no RNA) within condensates \cite{Hagan2023}. Our system consists of capsid subunits, cognate RNA, and non-cognate RNA with total concentrations $\rhot$, $\rhocogt$, and $\rhonct$, respectively. Subunits can adsorb to and subsequently assemble around both cognate and non-cognate RNA, forming RNA-filled capsids of size $N$ subunits. These components are immersed in a multicomponent solution which consists of a minority phase (the viral condensate) coexisting with a majority phase (the ``background''). We refer to the volumes of the condensate and background as $\Vc$ and $\Vbg$, which are related to the total system volume by $\VT=\Vc+\Vbg$. We present results in terms of the condensate size ratio (volume fraction),
$\Vr \equiv \Vc/\Vbg$. We showed in our previous work~\cite{Hagan2023} that assembly yields are typically maximized when $\Vr\ll1$, and the same is true of the model considered here (see Fig. S1). Thus, unless otherwise noted, we set $\Vr=10^{-3}$.
For simplicity we consider a single compartment, but our model is readily extended to multiple compartments and doing so does not qualitatively change the results. 

\subsection{Rate equation model}
\label{sec:rateEquationModel}

\subsubsection{Partition coefficients}
\label{sec:partionCoefficients}

Due to favorable interactions between viral components and condensate scaffold molecules, the compartment is enriched in the subunits and cognate RNA. This effect is quantified by partition coefficients $\Kc$, $\Kcog$, and $\KNC$ for subunits (i.e. capsid proteins), cognate RNA, and non-cognate RNA molecules, which set the \textit{equilibrium} ratios of concentrations in the condensate and background
\begin{align}
& \Kc =  \rhoonec/\rhoonebg \nonumber \\
& \Kcog =   \rhonma{cog}{0}{0}{\text{c}} /  \rhonma{cog}{0}{0}{\text{bg}} \nonumber \\
& \KNC =    \rhonma{nc}{0}{0}{\text{c}} /  \rhonma{nc}{0}{0}{\text{bg}}.
\label{eq:partitionCoef}
\end{align}
Here,  $\rho_1^{\alpha}$ is the concentration of monomers in phase $\alpha=\{\mbox{c, bg}\}$ and $\rhonma{$\nu$}{n}{m}{\alpha}$ is the concentration of RNA of species $\nu=\{\mbox{cog, NC}\}$ in phase $\alpha$ with $n$ adsorbed and $m$ assembled subunits. We emphasize that in Eq.~\ref{eq:partitionCoef} all concentrations are at their \textit{equilibrium} values. The partition coefficient $K_\nu$ is related to the change in solvation free energy $\Delta g_\nu$ for a molecule $\nu$ that transitions from the background to the condensate as $K_\nu = e^{-\beta \Delta g_\nu}$ with $\beta=1/\kt$ with $\kt$ the thermal energy. For simplicity, we assume that solvation interactions are additive and thus partition coefficients are multiplicative; i.e., the partition coefficient for a cognate RNA with $n$ adsorbed and $m$ assembled subunits is given by
\begin{align}
\Kcog(n,m) = &  \Kcog \Kc^{(n+m)}.
\label{eq:partitionCoefIntermediate}
\end{align}

\subsubsection{Capsid assembly model}
\label{sec:capsidModel}

To describe the assembly kinetics we adapt a rate equation formalism originally developed by Zlotnick \cite{Zlotnick1994,Zlotnick1999,Zlotnick2000, Endres2002, Hagan2010,Hagan2014, Morozov2009} to describe the 2-D polymerization of shells (capsids) in bulk solution. We have previously used versions of this model to describe condensate-coupled assembly of empty capsids \cite{Hagan2023} and capsids assembling around nanoparticles in bulk solution \cite{Hagan2009}. We briefly summarize the rate equations here and we comprehensively describe them in Appendix~\ref{app_sec:rate_equations}. The rate equations describe the time evolution of subunit and RNA concentrations.  We assume that subunits adsorb from solution (in background or condensate) onto the RNA with rate constant $\kads$ and free energy $\gads$ due to nonspecific electrostatic interactions between positive charges on capsid proteins and negative charges on the RNA backbone; the unbinding rate constant $\kads e^{\beta \gads}$ follows from detailed balance. The adsorbed subunits then undergo assembly into capsids of size $N$. An RNA molecule with $n$ adsorbed subunits and $m$ assembled subunits undergoes dimerization (if $m=0$) or association of a subunit to an assembly intermediate (if $m\geq 2)$ with rate $f_{n,m}$; disassembly occurs with rate $b_{n,m}$. Subunits and RNA molecules (with adsorbed and assembled subunits) can diffuse between the condensate and background; the associated rate constants are $\kdls$ and $\kdlr$, respectively. For the results in Fig.~\ref{fig:translation}, subunits are synthesized from cognate RNA with rate constant $\ktrans$; in all other main text figures, $\ktrans=0$. 

\subsection{Yields and selectivities}

We report two measures of assembly. (1) We define the total \emph{yield} $\yield$  as the fraction of RNA molecules in complete capsids, including cognate and non-cognate in the condensate and the background. Noting that the concentration of RNA molecules in capsids is equal to the concentration of complete capsids $\rhon$, 
\begin{equation}
    \yield = \rhon/\rhoRNA, \label{eq:yield}
\end{equation}
where $\rhoRNA=\rhocogt+\rhonct$ is the total RNA concentration. We analogously define the cognate yield $\yieldcog$ and non-cognate yield $\yieldnc$ as:
\begin{align}
    \yieldcog &= \rhoncog/\rhocogt \nonumber\\
    \yieldnc &= \rhonnc/\rhonct, \label{eq:yield_cog_nc}
\end{align}
with $\rhoncog$ and $\rhonnc$ the concentrations of capsids enclosing cognate and non-cognate RNA respectively. Yields of either species in the condensate and background are defined similarly (see Appendix~\ref{app_sec:bg_c_yields}). 

(2) We define the \emph{selectivity} as $s = \rhoncog/\rhonnc = \rhocogt\yieldcog/\rhonct\yieldnc $; we then define the \emph{normalized selectivity}, $\selecnorm$, as
\begin{equation}
    \selecnorm = \frac{2}{1+\yieldnc/\yieldcog}-1 \label{eq:selec_def}
\end{equation}
 such that $-1\leq \selecnorm \leq 1$. When $\selecnorm=1$, only cognate RNA is encapsidated; when $\selecnorm=-1$, only non-cognate RNA is encapsidated; when $\selecnorm=0$, cognate and non-cognate RNAs are encapsidated to an equal extent relative to the total concentration of each species. That is, when $\selecnorm=0$, $\rhoncog/\rhonnc = \rhocogt/\rhonct$.

\subsection{Equilibrium assembly theory}
\label{sec:equilibriumAssemblyTheory}

We also develop an equilibrium theory for our system, for three reasons. First, the equilibrium theory identifies conditions under which assembly is thermodynamically favorable, and the maximum (infinite-time) values of yield and selectivity. Second, we can derive approximate expressions that provide easily understandable relationships for how yields and selectivities depend on the key control parameters. Third, although in the long-time limit the rate equation results (with $\ktrans=0$) should closely match the equilibrium predictions (see Fig.~\ref{fig:yield_one_species}), at finite times slow nucleation or kinetic traps may cause the rate equation results to deviate significantly from equilibrium.

We calculate the equilibrium yields and selectivities using mass conservation, the law of mass action, and the definitions of the partition coefficients (Eqs. \ref{eq:partitionCoef} and \ref{eq:partitionCoefIntermediate}). In SI Section S2, we use these relationships to derive self-consistent equations for the equilibrium values of the yields (Eqs. \ref{eq:yield}, \ref{eq:yield_cog_nc}, and \ref{eq:yield_cog_nc_c_bg}). While these equations are complicated, we obtain simpler expressions for the equilibrium yield and selectivity in the limit of excess RNA ($\rhoRNA \gg \rhot/N$):
\begin{align}  
  \yield \approx
    \begin{cases}
        \frac{\rhot}{N\rhoRNA} (\cac/\rhot)^N, & \mbox{  for } \rhot\ll \cac  \\
        \frac{\rhot}{N\rhoRNA} (1-\cac/\rhot), & \mbox{  for } \rhot\gg \cac 
    \end{cases} \label{eq:equil_yield_approx} 
\end{align}
and
\begin{align}  \label{eq:equil_selec_approx}
    \selec &\approx \frac{\rho_{\text{cog},T} (1+\Vr \KNC)(1+\Vr \Kcog \Kc^N)}{\rho_{\text{NC},T}(1+\Vr \Kcog)(1+\Vr \KNC \Kc^N)}.
\end{align}
Here, $\cac$ is the critical assembly concentration:
\begin{align*}
    \cac &\approx e^{\beta(\gads+ \Gn/N)} \frac{1+\Vr\Kc}{1+\Vr} \\
    &\times\left(\sum_{\nu=\text{cog,NC}}\frac{1+\Vr K_{\nu} \Kc^N}{1+\Vr K_{\nu}}\rhonut\right)^{-\frac{1}{N}} \numberthis \label{eq:cac}
\end{align*}
Where the term $e^{\beta(\gads + \Gn/N)}$ (with $\Gn$ the free energy of an assembled capsid) reflects the subunit-RNA and subunit-subunit interactions, and the remaining terms account for translational entropy costs and interactions of RNA and subunits with the condensate.

\subsubsection{Typical parameter ranges}In typical  experiments, the condensate volume will be small compared to the background, $\Vr \ll 1$. For example, rotavirus viroplasms vary from 0.1$\upmu$m-$5\upmu$m in diameter \cite{Papa2021}, which gives $\Vr\approx10^{-7}-10^{-2}$ for a typical mammalian cell radius of $10\upmu$m. The typical total concentration of mRNA molecules in mammalian cells is on the order of $0.1\mu$M (BNID 109916 ~\cite{Milo2010}), while viral RNA concentrations can vary from $\approx$ 10 nM to a few $\upmu$M ~\cite{Borodavka2025}). These quantities can be varied over much larger ranges with \textit{in vitro} experiments. 

For convenience, we define all model parameters and give their values in Table~\ref{table:parameters} in the Appendix.

% Results and Discussion can be combined.
\section{Results}
\subsection{LLPS promotes robust encapsidation}\label{sec:robust}

\begin{figure}%[ht]
    %\centering
    \includegraphics[width=\linewidth]{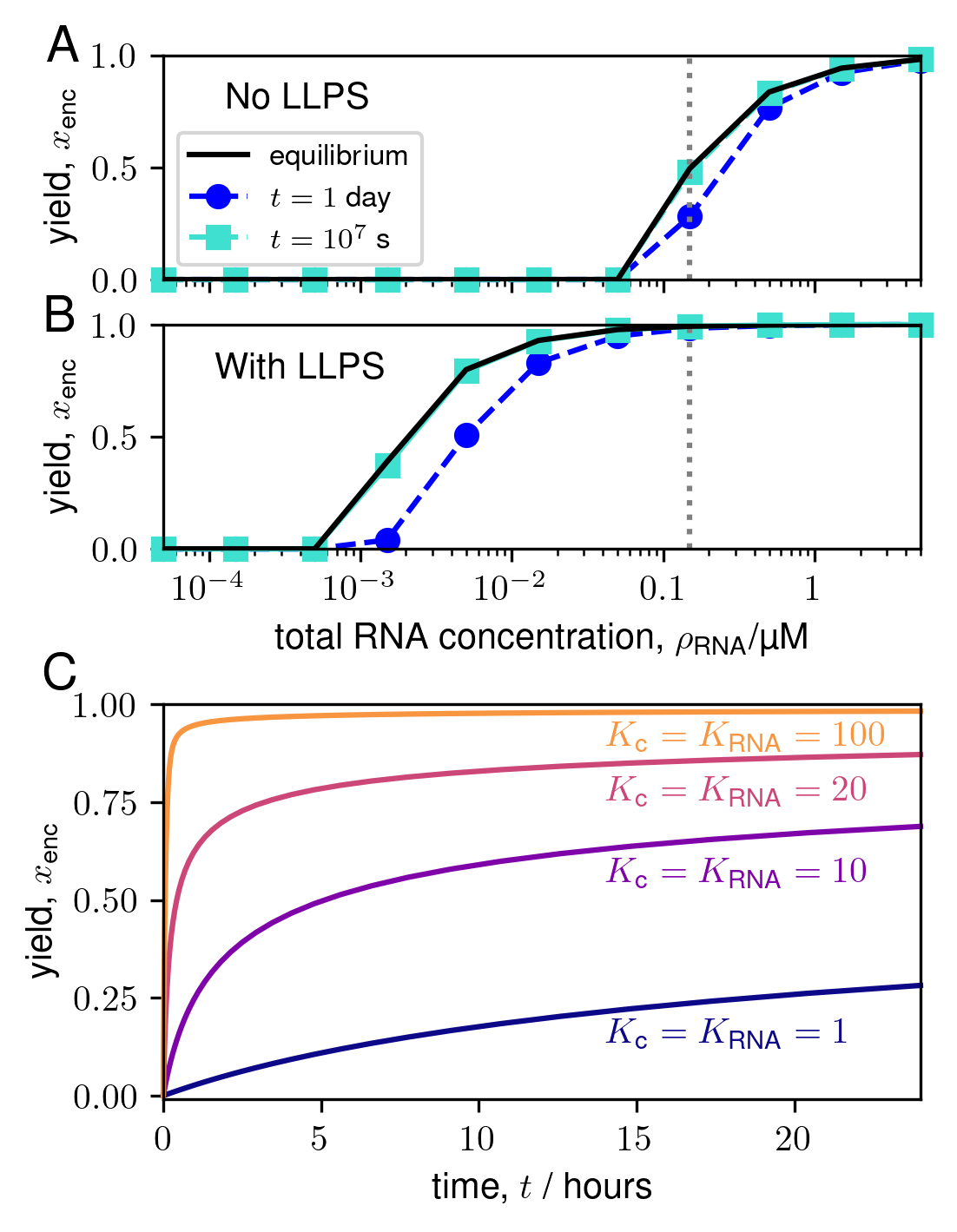}
     
   \caption{ \textbf{LLPS increases the robustness and rate of genome packaging .} This figure shows results for one RNA species (cognate). (A, B) The yield ($\yield$), defined as the fraction of RNA molecules in complete capsids (Eq.~\eqref{eq:yield}) as a function of RNA concentration $\rhoRNA$ without (A) and with (B) LLPS, for the subunit partition coefficient $\Kc=100$ and the RNA partition coefficient $\KRNA=100$. The total subunit concentration is $\rhot=N\rho_{\text{RNA}}$, so  there are just enough subunits to encapsidate every RNA molecule.
The dashed black lines are the equilibrium yield (Section S2B), the blue lines with circular markers are results from the rate equation model (Eq.~\ref{eq:rate_equation}) after $t=$1 day, and the turquoise lines with square markers are results from the rate equation model after $t=10^7$ seconds. The dotted gray vertical lines mark the RNA concentration ($0.15 \mu$M, close to a typical physiological RNA concentration of $\approx 0.1 \mu$M (BNID 109916 ~\cite{Milo2010})) at which the results in panel C were obtained.
   (C) Yield as a function of time from the rate equation model, for indicated values of the subunit and RNA partition coefficients.  
  }
    \label{fig:yield_one_species}
\end{figure}

We first investigate how LLPS affects assembly when there is only one species of RNA, and subunits and the RNA preferentially partition into the condensate, according to the partition coefficients $\Kc=\rhoonec/\rhoonebg$ and $\KRNA=\rhoRNAc/\rhoRNAbg$ (see the Model section). We consider a range of RNA concentrations, and for simplicity,  a constant stoichiometric ratio of subunits to RNA, $\rhot=N\rhoRNA$, so that there are just enough subunits to encapsulate all RNA molecules. Both the equilibrium and rate equation results show that the condensate increases the range of concentrations leading to productive assembly by orders of magnitude.  Fig.~\ref{fig:yield_one_species}A shows the yield (the fraction of RNA molecules in complete capsids, $\yield$) as a function of RNA concentration  without the effect of LLPS (i.e., no partitioning, $\Kc=1$, $\KRNA=1$), while Fig.~\ref{fig:yield_one_species}B shows yields with strong partitioning ($\Kc=100$, $\KRNA=100$). The equilibrium theory predicts that above a critical concentration $\cac$ the yield monotonically increases with RNA concentration until reaching nearly 100\%. While the rate equation yields are somewhat lower at the finite time of 1 day, they match the equilibrium results by long times ($10^7$ s). LLPS increases the range of concentrations at which assembly occurs by two orders of magnitude, by increasing the local concentration (i.e., within the condensate) by a factor of $\Kc=\KRNA=100$, which lowers $\cac$ by a factor of $\approx 100$ (see Eq.~\ref{eq:cac}). 

Additionally, the rate equation predicts that the condensate vastly increases assembly rates. Fig.~\ref{fig:yield_one_species}C shows yields as a function of time, demonstrating that both the assembly rate and the final yield dramatically and monotonically increase with $\Kc$ and $\KRNA$. These kinetics are shown for an RNA concentration of $\rhoRNA=0.15\mu$M (indicated by the gray dotted lines in panels A and B), which is roughly consistent with typical RNA concentrations in mammalian cells (BNID 109916 ~\cite{Milo2010}; ~\cite{Shapiro2013}). Thus, partitioning subunits and RNA into condensates enables rapid, high-yield encapsidation over a wide range of concentrations.

Note that the monotonic increase in yields at finite time with concentration contrasts with previous studies on empty capsid assembly, which find that the yield decreases at high concentrations due to a kinetic trap in which overly rapid capsid nucleation starves the system of free subunits before assembly can complete. This trap does not occur for our results at the stoichiometric ratio $\rhot=N\rhoRNA$ because the amount of nucleation is limited by the number of RNA molecules, as was shown previously for assembly of capsids around nanoparticles \cite{Hagan2009}. However, at sufficiently high concentrations, empty capsid assembly and malformed assemblies \cite{Hagan2009, Perlmutter2013, Perlmutter2014, Perlmutter2015}, which we have neglected in this study, would deplete finite-time yields. 

\subsection{LLPS provides a mechanism for selective viral genome packaging}\label{sec:selective_packaging}

\begin{figure}%[ht]
    %\centering
    \includegraphics[width=\linewidth]{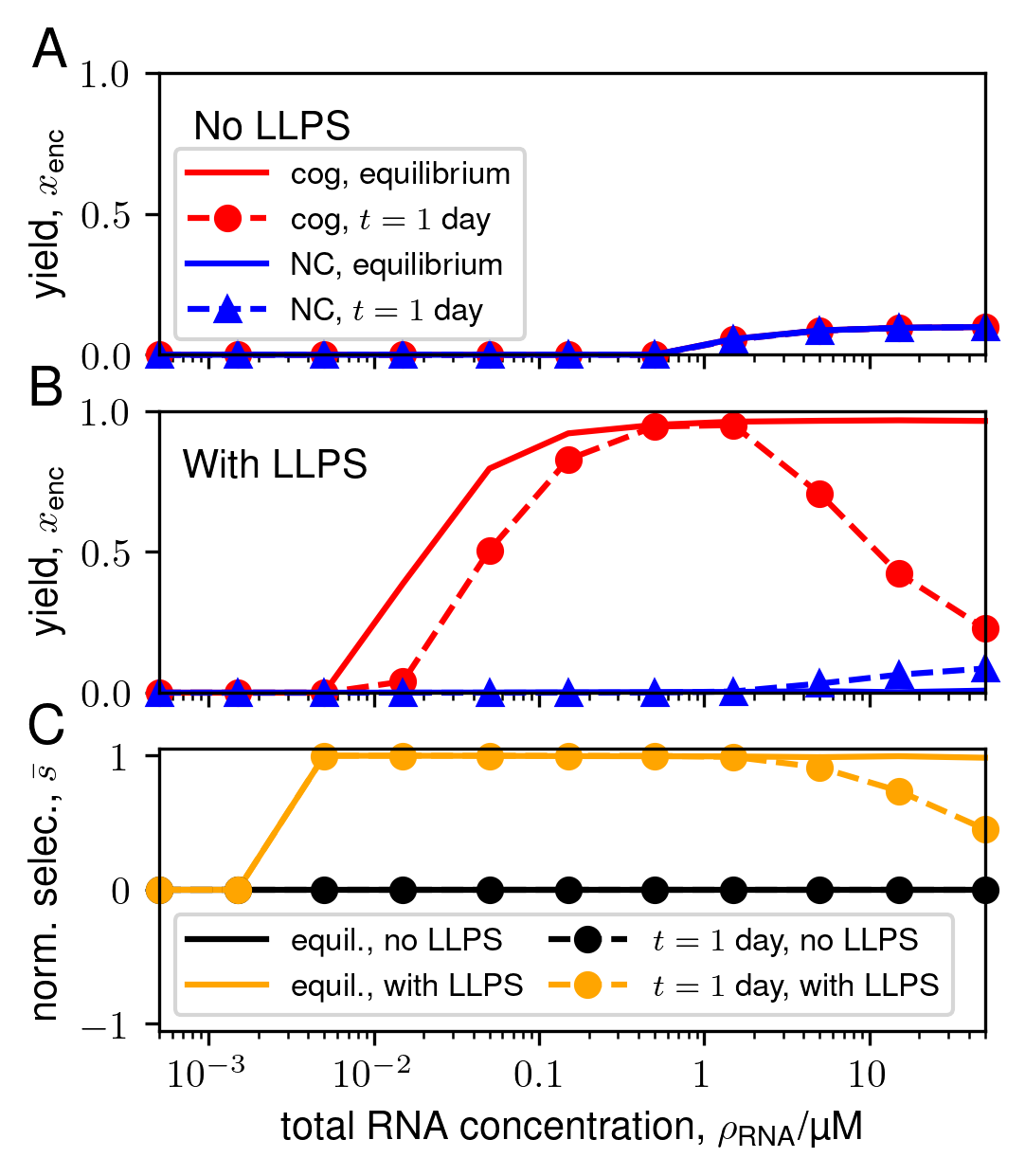}
    \caption{\textbf{LLPS drives selective packaging of cognate RNA over a wide range of concentrations.} (A, B) Yields of cognate (cog, red, circles) and non-cognate (NC, blue, triangles) RNA versus total subunit concentration, without (A) and with (B) LLPS. Solid lines denote equilibrium results (Section S2B), while dashed lines with markers correspond to rate equation results (Eq.~\ref{eq:rate_equation}) at $t=1$ day. (C) Normalized selectivity (Eq.~\ref{eq:selec_def}) with (orange) and without (black) LLPS. Solid lines denote equilibrium results, while dashed lines with markers correspond to rate equation results at $t=1$ day. For (A) and for black lines in (C), $\Kc=\Kcog=\KNC=1$, while for (B) and for orange lines in (C), $\Kc=\Kcog=100$ and $\KNC=10^{-2}$. For these results $\fcog=\rho_{\text{RNA,cog}}/(\rho_{\text{RNA,cog}}+\rho_{\text{RNA,NC}})=1/10$ and $\rhot=N\rho_{\text{RNA,cog}}$, so that there are just enough subunits to encapsidate every cognate RNA strand.
    }
    \label{fig:yield_two_species}
\end{figure}

We now consider a system with two RNA species: cognate (viral) and non-cognate (host). Consistent with the early stages of viral infections, where viral RNA is scarce compared to host RNA, we focus on conditions where the fraction of total RNA that is cognate is relatively small, $\fcog=\rhocogt/(\rhocogt+\rhonct)=0.1$. 
In the main text results we consider a stoichiometric ratio at which there are just enough subunits to encapsulate all the cognate RNA, $\rhot= N \fcog \rhoRNA$, but we present results in SI Section S1 (Figs. S4 -- S6) for other values of $\fcog$ and $\rhot$, which we discuss later in this section.

Cognate RNA molecules favorably partition into the condensate (with partition coefficient $\Kcog$) due to favorable interactions with condensate components, while non-cognate RNA tends to be excluded from the condensate (with partition coefficient ($\KNC$). Our model is agnostic to the molecular details of how this partitioning occurs. The partitioning mechanism is currently unknown for norovirus, whose ssRNA genome and nonstructural protein condensate most closely reflect our model \cite{Kaundal2024}. However, in rotavirus condensates (which have dsRNA genomes), the protein NSP2, a major constituent of the viral condensate, binds viral RNA to drive it partitioning~\cite{Nichols2024}. We will extend our model to account for such dsRNA genomes in future work. Host RNA could be excluded by less specific effects, e.g. an entropic penalty~\cite{Grigorev2025,Kelley2025}.  While either selective incorporation of cognate RNA into condensates or exclusion of non-cognate RNA from condensates alone can promote selective partitioning, we find that selectivity is most robust when both mechanisms operate. Note however that binding of subunits to either species of RNA increases its partition coefficient (see Eq.~\ref{eq:partitionCoefIntermediate}).

\begin{figure*}%[ht]
    %\centering
    \includegraphics[width=0.75\linewidth]{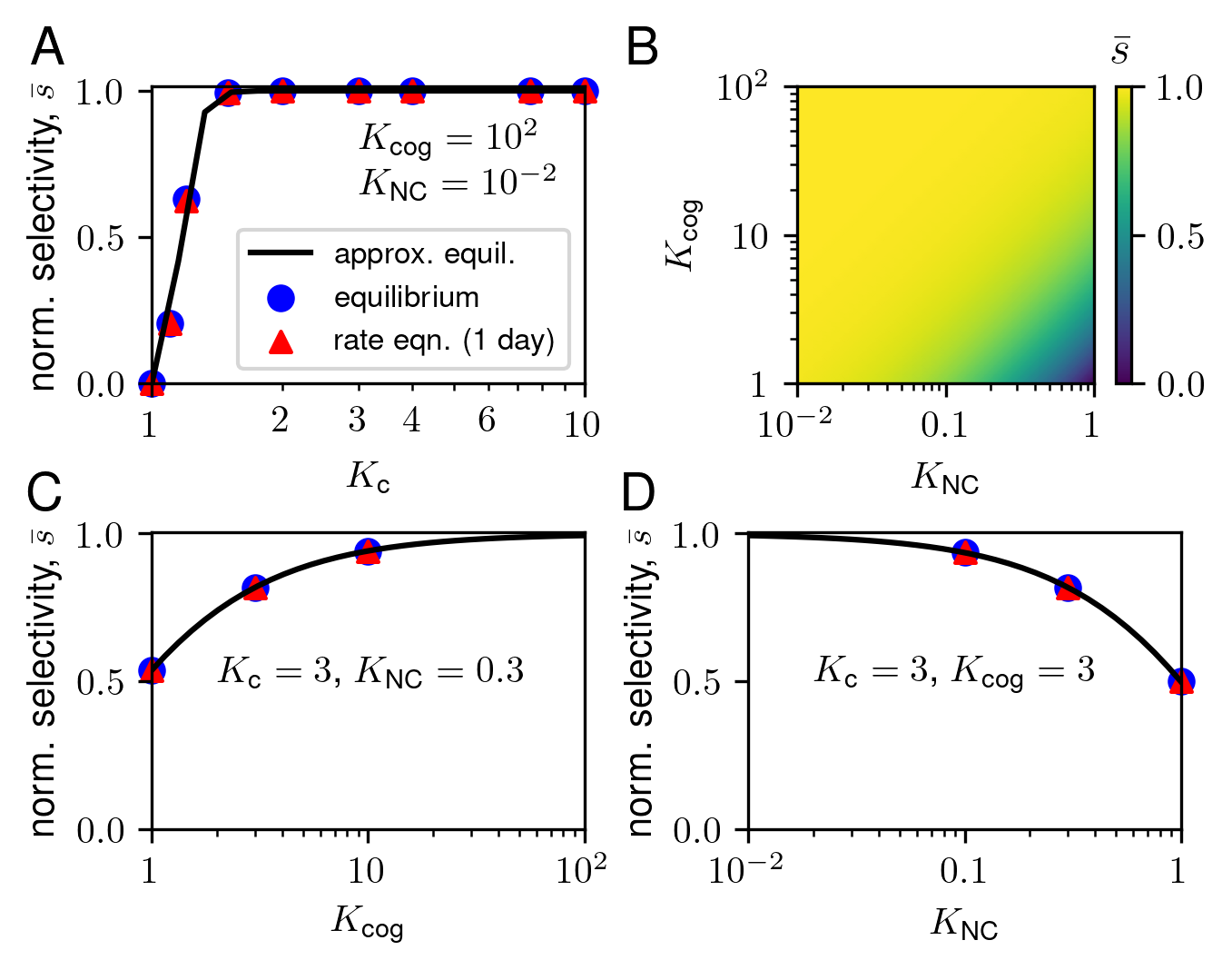}
    \caption{\textbf{LLPS promotes selective viral genome packaging over a wide range of partition coefficients.} (A) Normalized selectivity ($\selecnorm$, Eq.~\ref{eq:selec_def}) as a function of $\Kc$. Selectivity is shown for dynamics at $t=1$ day (red triangles), exact equilibrium (blue circles), and the approximate equilibrium result (solid black line, Eq.~\ref{eq:equil_selec_approx}), for $\rhonct=0.1\mu\text{M}$, $\fcog=0.1$, $K_{\text{cog}}=10^2$, and $K_{\text{NC}}=10^{-2}$. (B) Approximate equilibrium selectivity (Eq.~\ref{eq:equil_selec_approx}) as a function of  $K_{\text{cog}}$ and $K_{\text{NC}}$ for $\rhonct=0.1\mu\text{M}$, $\fcog=0.1$ and $\Kc=10^2$. (C) Selectivity as a function of $K_{\text{cog}}$ for $\rhonct=0.1\mu\text{M}$, $\fcog=0.1$, $\Kc=3$, and $K_{\text{NC}}=0.3$. (D) Selectivity as a function of $K_{\text{NC}}$ for $\rhonct=0.1\mu\text{M}$, $\fcog=0.1$, $\Kc=3$, $K_{\text{cog}}=3$. For all rate equation and exact equilibrium results in this figure, $\rhot=N\rhocogt$. Note that the approximate equilibrium result, Eq.~\ref{eq:equil_selec_approx}, is independent of $\rhot$.
    }
\label{fig:selectivity_figure}
\end{figure*}

Fig.~\ref{fig:yield_two_species}B,C show that these conditions  with  strong partitioning ($\Kc=\Kcog=100$ and $\KNC=10^{-2}$) drive essentially perfectly selective packaging of cognate RNA over a wide range of subunit concentrations. In particular, partitioning lowers the critical assembly concentration for cognate RNA by roughly two orders of magnitude, and above this threshold drives an equilibrium $\yieldcog$ approaching $100\%$, with  no assembly around noncognate RNA ($\selecnorm\approx 1$). In Fig. S2 we show how the cognate yield varies with time for several different values of the partition coefficients. As in the case of a single species of RNA (Fig.~\ref{fig:yield_one_species}C), partitioning increases both the assembly rate and the final ($t=1$ day) yield. For a wide range of $\rhoRNA$, the rate equation results nearly reach equilibrium by 1 day. However, at high RNA concentration $\rhoRNA \gtrsim  1 \upmu$M (an order of magnitude above typical physiological concentrations) the cognate yield drops markedly and non-cognate yield increases slightly. This difference between finite time and equilibrium reflects a kinetic trap: at such high concentrations the nucleation rate in the background becomes comparable to the rates of subunit diffusion into and assembly within the condensate (see Fig. S3, as well as SI Section S3 for timescale estimates). Consequently, subunits become ``stuck'' in partially- or fully-assembled capsids containing non-cognate RNA.  

 Fig.~\ref{fig:yield_two_species}A shows the control example with no LLPS effects ($\Kc=\Kcog=\KNC=1$). Here, there is no distinction between cognate and non-cognate RNA, and thus no selectivity ($\selecnorm=0$, black lines in Fig.~\ref{fig:yield_two_species}C) --- capsids assemble around RNA species in proportion to their concentrations. The overall yield (fraction of each RNA species enclosed, $\yield$) is limited to $\fcog=0.1$ by the subunit stoichiometric ratio  ($\rhot/ N \rhoRNA = \fcog$); higher stoichiometric ratios of subunits lead to higher yields (Fig. S4).

\textit{Varying stoichiometric ratio and fraction cognate.}
The results thus far consider a limiting subunit stoichiometric ratio $\rhot=N\rhocogt$ and one value of $\fcog=0.1$.
Fig. S4 shows results for a higher subunit stoichiometric ratio $\rhot=N\rhoRNA$, sufficient to  encapsidate \textit{all} RNA molecules.  LLPS still leads to selective encapsidation in this case, although over a significantly narrower range of concentrations compared to (Fig.~\ref{fig:yield_two_species}B).
Fig. S5 shows  that, for fixed $\rhonct$, the yields decrease with decreasing $\fcog$, consistent with the simplified equilibrium expression Eq.~\ref{eq:equil_yield_approx}, due to the increasing chemical potential cost to remove a cognate RNA from solution as its concentration decreases. Importantly though, selectivity remains high throughout the range over which cognate RNA is encapsidated.  
Moreover, high yields can be achieved even at low $\fcog$ by increasing the total RNA concentration $\rhoRNA$ and thus the cognate concentration (Fig. S6).

\textit{Varying partition coefficients.}
Fig.~\ref{fig:selectivity_figure} shows how selectivity depends on the partition coefficients for the typical non-cognate RNA concentration $\rhoRNA=0.1\upmu$M.
Fig.~\ref{fig:selectivity_figure}A shows varying subunit partition coefficient $\Kc$, demonstrating that selectivity remains extremely high even for low $\Kc$ under strong partitioning/exclusion of cognate/non-cognate RNA. For $\Kc\gtrsim2$, capsids assemble almost exclusively around cognate RNA ($\bar{s}\approx 1$). Note that the approximate equilibrium selectivity (Eq.~\ref{eq:equil_selec_approx}), which depends only on the ratio of total cognate and non-cognate concentrations and not on the concentrations themselves, closely matches the exact equilibrium result and the rate equation results at $t=1$ day. Thus, Eq.~\ref{eq:equil_selec_approx} serves as a simple and accurate predictor of how much partitioning is required to achieve a desired selectivity. Fig.~\ref{fig:selectivity_figure}B shows Eq.~\ref{eq:equil_selec_approx} as a function of $\Kcog$ and $\KNC$ for $\Kc=100$, demonstrating that when subunit partitioning is strong, only modest partitioning of cognate into, and/or non-cognate out of, the condensate is required to achieve high selectivity.  Figs.~\ref{fig:selectivity_figure}C, D show how the selectivity depends on $\Kcog$ and $\KNC$ for weak subunit partitioning $\Kc=3$ and non-cognate (C) or cognate (D) RNA partitioning. Selectivities close to one are attained for $\Kcog \gtrsim 10$ (C) and $\KNC \lesssim 0.1$ (D). Taken together, these results show that LLPS enables highly selective packaging of cognate RNA at physiologically realistic concentrations and partition coefficients.

\subsection{Subunit translation further enhances packaging selectivity}
\label{sec:translation}

The concentration of viral proteins is not constant in infected cells, since subunits are synthesized from (positive sense) viral RNA by ribosomes. Some viruses incorporate ribosomes within viroplasms \cite{Desmet2014,Tenorio2019,FernandezdeCastro2014,Wang2022,Liu2024,Jobe2024}, suggesting the possibility of protein translation within condensates. Therefore, we now consider the consequences of subunit translation for condensate-mediated selective packaging. We assume that subunit translation occurs at a rate proportional to local cognate RNA concentration (see Model section). However, since the actual rate will depend on ribosome concentration along with other factors, we treat the translation rate constant $\ktrans$ as a variable parameter.

\begin{figure}%[ht]
    %\centering
    \includegraphics[width=\linewidth]{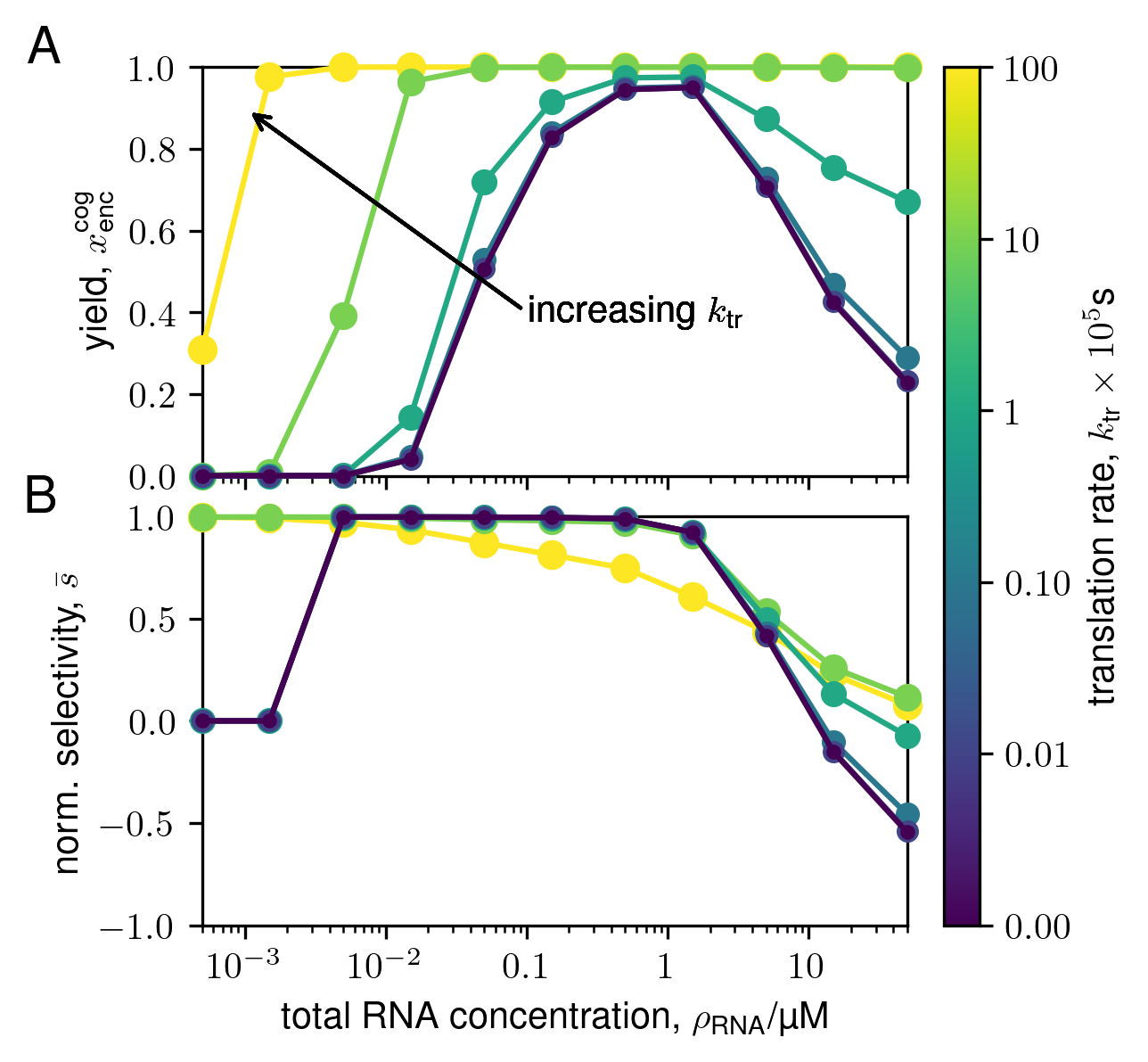}
    \caption{\textbf{Translation of capsid proteins in the condensate increases selective packaging of the viral genome.} (A) Fraction of cognate RNA in complete capsids ($\yieldcog$) as a function of total RNA concentration for indicated capsid protein translation rates. (B) Normalized selectivity $\selecnorm$ for cognate RNA as a function of total RNA concentration for indicated translation rates. Other parameters are $\Kc=\Kcog=10^2$, $\KNC=10^{-2}$, $\fcog=0.1$, $\Vr=10^{-3}$. For $\ktrans=0$, $\rhot=N\rhocogt$, so that there are just enough subunits to encapsidate all cognate RNA strands.}
    \label{fig:translation}
\end{figure}

 Fig.~\ref{fig:translation}A,B show how the yield and selectivity vary with the total RNA concentration for a range of translation rates. 
Fig.~\ref{fig:translation}A shows that increasing translation rates makes assembly significantly more robust, increasing the range of total RNA concentrations that result in high cognate yields by up to 100-fold. Fig.~\ref{fig:translation}B shows that the effect of translation on selectivity is more nuanced at these parameters. Translation enables high selectivity at  low RNA concentrations ($\rhoRNA \lesssim 0.01 \upmu$M, where no assembly occurs wth $\ktrans=0$), and modestly increases the selectivity at very high RNA concentrations ($\rhoRNA \gtrsim 1 \upmu$M). The latter effect arises because subunits are preferentially being synthesized inside the condensate, circumventing the ``monomer starvation'' kinetic trap due to subunits binding to non-cognate RNA in the background. However, since  selectivity is already 100\%  over the range $\rhoRNA \in (0.01, 1) \upmu$M at these parameters, translation has little effect in this range except for slightly decreasing selectivity at extremely high rates ($\ktrans=10^{-3}\text{s}^{-1}$). This effect arises because high translation rates lead to high subunit stoichiometric ratios (in both the condensate and background), causing capsids to assemble around non-cognate and cognate RNA.

\section{Discussion}
Our results show that coupling self-assembly to biomolecular condensates is a powerful strategy for making viral genome packaging robust, rapid, and selective. Despite the significant excess non-cognate RNA present in an infected cell, preferential partitioning of subunits and cognate RNA into the condensate leads to high yields of capsids containing viral genomes and high selectivity --- low yields of capsids with non-cognate RNA. For physically relevant ranges of partition coefficients, we observe essentially perfect selectivity over wide ranges of RNA and capsid protein concentrations. Although selective packaging does not require exclusion of host genetic material from the condensate, selectivity is optimized when viral condensates both enrich themselves with viral structural components and exclude host genetic material. We also find that partitioning of subunits and cognate RNA into the condensate can increase assembly rates and robustness (the range of concentrations leading to high yields) by orders of magnitude, consistent with previous studies of empty capsid assembly~\cite{Hagan2023,Frechette2025}. 

Motivated by the observation of ribosomes in or near some viral condensates, our model predicts that protein translation can enhance selective packaging. While previous work has shown that assembly can be improved by addition of capsid proteins during assembly \cite{Hagan2011, Castelnovo2014, Dykeman2014}, a key difference in our system that enhances selectivity is that translation is proportional to the local viral RNA concentration, which is higher in the condensate.

\textit{Comparison with packaging signals and other mechanisms for selective packaging.} While several mechanisms have been proposed for selective viral packaging \cite{Comas-Garcia2019}, most work has focused on packaging signals, which are specific RNA sequences that interact with sites on capsid proteins. Here, we show that an alternative mechanism, co-recruitment of capsid proteins and viral RNA molecules into a condensate, is also a powerful mechanism to drive selectivity.  While for simplicity we did not consider packaging signals in this model, it seems likely that selectivity could be further enhanced by combining packaging signals with condensate-coupled assembly. Thus, a natural next step will be to incorporate packaging signals into our model.

Indeed, a number of viruses that employ condensates during genome packaging are known to have packaging signals, such as HIV \cite{Keane2015}, rotavirus \cite{McDonald2011}, and reovirus \cite{Roner2007}. It is also worth noting that the viral condensate scaffolds must have a mechanism to distinguish viral from host RNA, which likely involves sequence-specific interactions; for example, rotavirus RNA is recruited by the viral RNA binding protein NSP2. Alternatively, favorable capsid protein-condensate interactions among capsid proteins adsorbed to viral RNA could facilitate partitioning of viral RNA (as observed for SARS-CoV-2~\cite{Jack2021}). Finally, non-cognate RNA, which does not have such preferential interactions with viral proteins, could be preferentially excluded from the condensate by generic entropic forces ~\cite{Grigorev2025,Kelley2025}.

\textit{Testing in experiments.} 
Because our models are general and do not depend on specific molecular details, their predictions can apply to diverse experimental systems in which assembly subunits and a cargo preferentially partition into phase-separated condensates. As noted in the introduction, there are cellular systems in which multiple components preferentially partition into condensates to promote co-assembly, such as clathrin cages and cargo during exocytosis \cite{Ravindran2024,Dragwidge2024,Kozak2022,Bergeron-Sandoval2021,Mondal2022,Wilfling2023}. Selective partitioning and exclusion of different molecules is a common feature among biomolecular condensates~\cite{Banani2017,Ditlev2018,Lyon2021,Grigorev2025,Kelley2025,AmbadiThody2024}, and thus likely represents a general strategy for promoting desired reactions while simultaneously preventing undesirable ones. Because selectivity depends on $\Kc$ as $s\propto \Kc^N$ (see Eq.~\ref{eq:equil_selec_approx}), this strategy is particularly advantageous for high-order reactions (such as viral genome packaging). 

The biological systems that most directly motivate this work are viral condensates in which capsids assemble around the genome (e.g. \cite{Kaundal2024,Geiger2021,He2024}). Our theoretical predictions could be tested in these systems if there is a way to manipulate partition coefficients of viral RNAs and capsid proteins, or by controlling other variables such as RNA concentration. 

However, it would be more straightforward to directly test our predictions with  \textit{in vitro} experiments on phase separated-condensates, which enable more direct control over concentrations and partition coefficients. For example, the capsid proteins and RNA from several viral systems have been shown to form condensates in vitro \cite{Cubuk2021,Geiger2021,Guseva2020,He2024}. More broadly, DNA origami and protein design have recently been used to engineer subunits that assemble into capsids \cite{Sigl2021,Wei2024,Divine2021,Butterfield2017,Bale2016,Hsia2016,Ren2019,Laniado2021,McConnell2020}. In the DNA origami system, the capsids were further engineered to assemble around cargoes of DNA or functionalized nanoparticles. Our predictions could be tested by engineering such subunits to favorably partition into the condensate (e.g. through DNA `handles' or other types of interactions), and two species of cargoes that have been functionalized to have different partition coefficients for the condensate. 

\textit{Outlook.}
Given that subunit translation can enhance selective packaging, it would be interesting to extend our model to consider the timing of translation, since previous computational studies suggest that continuously modulating the concentration of available subunits can optimize self-assembly~\cite{Jhaveri2024, Dykeman2014}. Further, to simplify a large parameter space, we have assumed a fixed RNA concentration in this work, but the model can be extended to account for RNA transcription. Finally, it will be important to test model assumptions, such as the influence of excluded volume \cite{Frechette2025} and RNA structure \cite{Perlmutter2013, Li2018, Li2022a, Panahandeh2022, Safdari2025}, using particle-based molecular dynamics simulations. 

Such studies will advance our understanding of condensate-mediated assembly as a fundamental mechanism for selective co-assembly in biology. They also have practical implications for biomedicine and nanomaterials. Understanding LLPS could lead to new antiviral therapy approaches that block assembly and/or selective genome packaging, and could guide the development of condensates as a powerful approach for nanoscale packaging in human-engineered systems. 

\begin{acknowledgments}
We thank Mauricio Comas-Garcia and Adam Zlotnick for helpful comments on an earlier version of this manuscript. This work was supported by the NSF through DMR 2309635 and the Brandeis Center for Bioinspired Soft Materials, an NSF MRSEC (DMR-2011846). Computing resources were provided by the National Energy Research Scientific Computing Center (NERSC), a Department of Energy Office of Science User Facility (award BES-ERCAP0026774); the NSF ACCESS allocation TG-MCB090163; and the Brandeis HPCC which is partially supported by the NSF through DMR-MRSEC 2011846 and OAC-1920147.
\end{acknowledgments}

\section*{Data Availability}
Code for running simulations and for reproducing figures is available on GitHub at: [link to be inserted upon publication].

\appendix

\section{Rate Equations}\label{app_sec:rate_equations}
\begin{table*}
%\centering
\begin{tabular} { | c | c | c | }
  \hline
  Parameter & Value(s) & Description   \\
  \hline
  $\rhot$ ($\upmu$M) & $10^{-3}$--100 & total capsid subunit concentration \\ 
  $\rhocogt$ ($\upmu$M) & $5\times 10^{-5}$--5 & total cognate RNA concentration \\ 
  $\rhonct$ ($\upmu$M) & $4.5\times 10^{-4}$--45 & total non-cognate RNA concentration \\ 
  $\VT$ ($\upmu\text{m}^3$) & 4190 & total volume \\
  $\Vr$ & $10^{-3}$ & condensate size ratio \\
  $\Kc$ & 1--100 & subunit partition coefficient \\
  $\Kcog$ & 1--100 & cognate RNA partition coefficient \\
  $\KNC$ & 0.01--1 & non-cognate RNA partition coefficient \\
  $\gNuc$ ($\kt$) & -5 & subunit binding free energy in nucleation phase \\
  $\gElong$ ($\kt$) & -10 & subunit binding free energy in elongation phase \\
  $\gn$ ($\kt$) & -20 & binding free energy of final subunit \\
  $\gads$ ($\kt$) & -3.5 & adsorption free energy \\
  $\kads$ ($\upmu\text{M}^{-1}\text{s}^{-1}$) & $10^6$ & adsorption rate constant \\
  $N$ & 20 & number of subunits in complete capsid \\
  $\nnuc$ & 3 & critical nucleus size \\
  $\fcog$ & 0.1 & ratio of cognate RNA to total RNA \\
  %$\rpr$ & 1, 10 & stoichiometric ratio of subunits to RNA \\
  $D_1$ ($\upmu$m$^2$/s) & 10 & subunit diffusion constant \\ 
  $D_{\text{RNA}}$ ($\upmu$m$^2$/s) & 1 & RNA diffusion constant \\ 
  $\fElong$ (s$^{-1}$) & $10^4$ & assembly rate constant \\ 
  \hline
\end{tabular}
\caption{Simulation parameter values and descriptions.} %Energies are expressed in units of $\kt$ ($\approx 0.6$ kcal/mol at room temperature).} 
\label{table:parameters}
\end{table*}
To model the kinetics of LLPS-coupled genome packaging, we use the following rate equations:
\begin{subequations}
    \begin{align}
        \frac{\dee \rhoonea}{\dee t} &= \mathcal{N}_1^{\alpha} + \mathcal{D}_1^{\alpha} + \mathcal{T}^{\alpha} \\
        \frac{\dee \rhonma{$\nu$}{n}{m}{\alpha}}{\dee t} &= \adsraterna + \assemraterna + \diffraterna,
    \end{align} \label{eq:rate_equation}
\end{subequations}
where $\nu=\{\mbox{cog, NC}\}$ and $\alpha=\{\mbox{c,bg}\}$. The quantities $\adsratesub$, $\diffratesub$, and $\transratesub$ represent, respectively, the rate of change in free subunit concentration due to adsorption and desorption, the diffusive exchange of free subunits, and (when applicable) the rate of subunit translation from mRNA molecules:
\begin{widetext}
\begin{subequations}
    \begin{align}
        \adsratesub &= -\kads \rho_1^{\alpha} \sum_{m=0}^N{}^{'} \sum_{n=0}^{N-m} \left(N-(n+m)\right) \sum_{\nu=\mbox{cog, NC}}\rhonma{$\nu$}{n}{m}{\alpha} 
        + \kdes \sum_{m=0}^N{}^{'} \sum_{n=0}^{N-m} n \sum_{\nu=\mbox{cog, NC}}\rhonma{$\nu$}{n}{m}{\alpha} \\
        \diffratesub &= \begin{cases}
        \frac{1}{\Vc} \kdls\left(\rhoonebg - \rhoonec/\Kc\right), &\alpha=\text{c} \label{eq:subunit_diff}\\
        -\Vr \diffratesubc, &\alpha=\text{bg}
        \end{cases}\\
        \transratesub &= \ktrans \sum_{m=0}^{N-1}{}^{'} \sum_{n=0}^{N-m}\rhonma{cog}{n}{m}{\alpha}.
    \end{align}
\end{subequations}
\end{widetext}
%\mfh{Use `nameref' for section references.}
Here, $\kads$ is the rate constant for adsorption of subunits to RNA, and $\gads$ is the adsorption free energy; the desorption rate constant $\kdes = \cSS \kads e^{\beta \gads}$, with the standard state concentration $\cSS=1$M, follows from detailed balance. The sums over $n$ run over the number of adsorbed (but unassembled) subunits; the sums over $m$ run over the number of assembled subunits, and the prime symbol indicates that the sum excludes $m=1$ (there are no ``single'' assembled subunits); and the sums over $\nu$ run over $\{\mbox{cog, NC}\}$. For simplicity, we assume that a maximum of $N$ subunits can adsorb to a single RNA molecule. Thus, the sums over $n$ have an upper limit of $N-m$. The diffusion-limited rate constant $\kdls=4\pi\Rc D_1$ sets the speed at which subunits enter the condensate. The condensate radius is given by $\Rc=(\Vc/(4\pi/3))^{1/3}$, and we set $D_1=10\upmu$m$^2$/s (a typical value for proteins in the cytoplasm (BNID 112087 \cite{Milo2010}). Finally, $\ktrans$ is the translation rate constant. We consider the effects of translation in Section~\ref{sec:translation}; in other cases we set $\ktrans=0$. Note that, when we do include translation, we assume that RNA with adsorbed and partially assembled subunits is still accessible for translation. We have tested this assumption by using alternative models in which either (1) only RNA which has no adsorbed subunits or (2) all RNA except those which are fully encapsidated are accessible for translation, and we find very similar behavior for all three models (see Fig. S7).

The quantities $\adsraterna$, $\assemraterna$, and $\diffraterna$ are the rates of change of RNA concentrations due to adsorption, assembly, and RNA diffusion, respectively:
\begin{widetext}
\begin{subequations}
\begin{align}
    \adsraterna &= \kads \rhoonea\rhonma{$\nu$}{n-1}{m}{\alpha} - \kads \rhoonea\rhonma{$\nu$}{n}{m}{\alpha} 
    - \kdes \rhonma{$\nu$}{n}{m}{\alpha} + \kdes \rhonma{$\nu$}{n+1}{m}{\alpha}\\
    \assemraterna &= \begin{cases}
        b_{n-2,2}\rhonma{$\nu$}{n-2}{2}{\alpha}-f_{n+2,0}\rhonma{$\nu$}{n+2}{0}{\alpha}, &m=0 \\
        f_{n+2,0}\rhonma{$\nu$}{n+2}{0}{\alpha} - (f_{n,m}+b_{n,m})\rhonma{$\nu$}{n}{m}{\alpha} + b_{n-1,m+1}\rhonma{$\nu$}{n-1}{m+1}{\alpha}, &m=2 \\
        f_{n+1,m-1}\rhonma{$\nu$}{n+1}{m-1}{\alpha} - (f_{n,m}+b_{n,m})\rhonma{$\nu$}{n}{m}{\alpha} + b_{n-1,m+1}\rhonma{$\nu$}{n-1}{m+1}{\alpha},&m>2  \\
        f_{n+1,N-1}\rhonma{$\nu$}{n+1}{N-1}{\alpha} - b_{n,N}\rhonma{$\nu$}{n}{N}{\alpha}, &m=N
    \end{cases}\\
    \diffraternaopt{\nu}{\alpha} &= \begin{cases} \frac{1}{\Vc}\kdlr \left(\rhonma{$\nu$}{n}{m}{\text{c}}- \rhonma{$\nu$}{n}{m}{\text{bg}}/\left(K_{\nu}\Kc^{(n+m)}\right)\right), & \alpha=\text{c}\\
    -\Vr \diffraternaopt{\nu}{\text{c}}, &\alpha=\text{bg} 
    \end{cases} \label{eq:rna_diff}
\end{align}
\end{subequations}
\end{widetext}
The RNA diffusion-limited rate constant is $\kdlr=4\pi\Rc D_{\text{RNA}}$. We assume for simplicity that $D_{\text{RNA}}=1\upmu$m$^2$/s, independent of how many subunits are adsorbed; the lower value of $D_{\text{RNA}}$ compared to $D_1$ reflects the larger size of RNA compared to capsid proteins.

Molecular dynamics simulations have shown that capsid assembly on nucleic acids and other polymers involves a complex dynamics including cooperative polymer-subunit motions \cite{Elrad2010, Perlmutter2013, Perlmutter2014,Hu2007}. Importantly though, subunit association rates depended primarily on the concentration of adsorbed but unassembled subunits, with a weak dependence on polymer length \cite{Elrad2010}. Therefore, for the simple model in this work  we follow Dykeman et al.~\cite{Dykeman2013} and assume that intermediates can only associate with subunits at adjacent sites on the polymer. The forward assembly rates are then given by
\begin{subequations}
\begin{align}
    f_{n,m} &= \begin{cases}
        \fElong n/(N-m), &0<m<N \\
        \fElong n(n-1)/(2N), &m=0,\; n>0 \\
        0, &\text{otherwise} \label{eq:fnm} 
    \end{cases} 
\end{align}
Here the term $n/(N-m)$ reflects the probability that an adsorbed subunit is adjacent to the assembly intermediate; the additional factor $(n-1)/2$ in the second line of Eq.~\ref{eq:fnm} accounts for the fact that dimerization can happen anywhere along the RNA molecule. The term $\fElong$ is the association rate constant  given the presence of an adjacent subunit. We set $\fElong=10^4$s$^{-1}$ (see SI Section S4 and \cite{Dykeman2013} for an explanation of this estimate). 

The dissociation rate constants then follow from detailed balance as
\begin{align}
    b_{n,m} &= \begin{cases}
        \fElong e^{\beta g_m}, &m>0 \\
        0, &\text{otherwise} \label{eq:bnm}
    \end{cases}
\end{align}
\end{subequations}

\textbf{\textit{Assembly model.}}
Following Refs. \cite{Zlotnick1994,Zlotnick1999,Zlotnick2000, Endres2002, Hagan2010,Hagan2014, Morozov2009}, the assembly reaction is split into `nucleation' and `elongation'  (or `growth') phases, to reflect the fact that smaller intermediates have fewer attractive contacts per subunit and are thus less stable compared to larger intermediates. Therefore, below a critical nucleus size $\nnuc$, the excess free energy of subunit binding is unfavorable at productive reaction conditions whereas association in the elongation phase is favorable,  driving forward-biased assembly. We set $\nnuc=3$, $N=20$, $\gNuc=-5\kt$, $\gElong=2\gNuc$, and $\gn=4\gNuc$. The especially favorable value of $\gn$ reflects the fact that in most capsid structures, the final subunit makes more contacts than previous steps in the elongation phase. We assume that $\gads$ and the subunit-subunit affinities are the same in the condensate and background.

This model assumes a single simplified assembly pathway, with a single intermediate per RNA molecule. 
While the approach can be extended to allow for additional assembly pathways \cite{Endres2005}, this can be done more efficiently using stochastic simulations \cite{Dykeman2013, Dykeman2017, Schwartz1998, Smith2016}. Moreover, the spatial structure of assembly can be accounted for using particle-based molecular dynamics (MD) simulations (e.g. \cite{Hagan2006,Elrad2010,Perlmutter2013,Perlmutter2014,Perlmutter2015a,Frechette2025a,Timmermans2022,Panahandeh2022,Li2022a,Gupta2023,Nguyen2007,Nguyen2009}) or the reaction-diffusion framework NERDDS \cite{Varga2020}. Accounting for the effects of RNA on assembly pathways and intermediate structures will be most straightforward to accomplish using MD simulations (eg. \cite{Perlmutter2013, Perlmutter2014, Perlmutter2015, Panahandeh2022, Li2022a, Safdari2025, Mahalik2012, Zhang2014}).
We have also assumed that association of subunits in solution to intermediates occurs at negligible rates in comparison to RNA-bound subunits. Furthermore, we neglect the possibility of empty capsids since these will be present at extremely low populations in comparison to RNA-filled capsids for the parameters we study: the relative equilibrium concentrations of empty and filled capsids is $\frac{\rhon(\text{filled})}{\rhon(\text{empty})} = \rhorf \exp( -\beta N \gads) \ll 1 $  with $\rhorf$ the free RNA concentration \cite{Hagan2009,Zandi2009}. Consistent with this prediction, most viruses that use LLPS do not assemble empty virions.

\section{Yields in condensate and background} \label{app_sec:bg_c_yields}
Condensate cognate/non-cognate and background cognate/non-cognate yields are defined as: 
\begin{align}
    \yieldCnu &= \frac{\rhonnuc}{\rhonut}\frac{\Vr}{1+\Vr} \nonumber \\
    \yieldBGnu &= \frac{\rhonnubg}{\rhonut}\frac{1}{1+\Vr}, \label{eq:yield_cog_nc_c_bg}
\end{align}
where $\rhonnua$ is the concentration of capsids enclosing RNA of species $\nu=\{\mbox{cog, NC}\}$ in phase $\alpha=\{\mbox{c,bg}\}$. Fig. S3 shows $\yieldCnu$ and $\yieldBGnu$ versus time for moderate and high RNA concentrations. At moderate RNA concentrations, there is no assembly in the background at any time; at high RNA concentrations there is a small yield in the background at short times, but these capsids quickly diffuse into the condensate.

\bibliography{RNA_LLPS_references}

\end{document}

\title{Supplementary Information: Liquid-liquid phase separation promotes selective viral genome packaging}

\author{Layne B. Frechette}
\affiliation{
 Martin Fisher School of Physics, Brandeis University, Waltham, Massachusetts 02453, USA
}

\author{Michael F. Hagan}
\email{hagan@brandeis.edu}
\affiliation{
 Martin Fisher School of Physics, Brandeis University, Waltham, Massachusetts 02453, USA
}

\date{\today}

\maketitle
\onecolumngrid
\tableofcontents

\section{Additional Figures}
\label{supsect:figures}

\begin{figure}%[ht]
    \centering
    \includegraphics[width=0.5\linewidth]{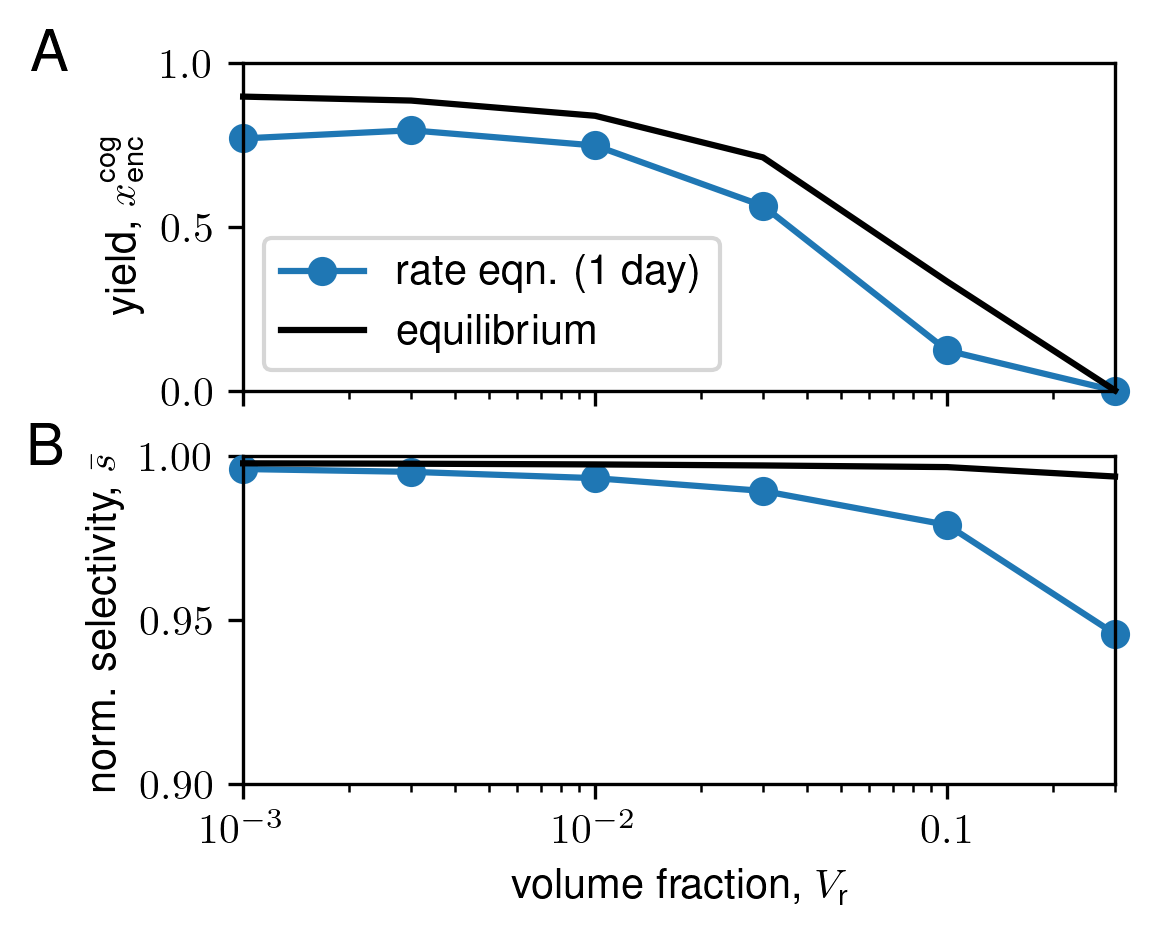}
   \caption{ \textbf{Lowering the condensate volume fraction increases yield and selectivity.}
   (A) Yield and (B) selectivity as a function of $\Vr$, with $\Kc=\Kcog=100$, $\KNC=10^{-2}$, $\fcog=0.1$, and $\rhonct=0.1\mu$M. Here $\rhot=N\rho_{\text{RNA,cog}}$, so that there are just enough subunits to encapsidate every cognate RNA strand. The black lines show full equilibrium results (Section \ref{supp_sec:equil_two_species}), while the blue lines with circles show rate equation results at $t=1$ day.
   }
    \label{supp_fig:yield_selectivity_vary_vr}
\end{figure}

\begin{figure}%[ht]
    \centering
    \includegraphics[width=0.5\linewidth]{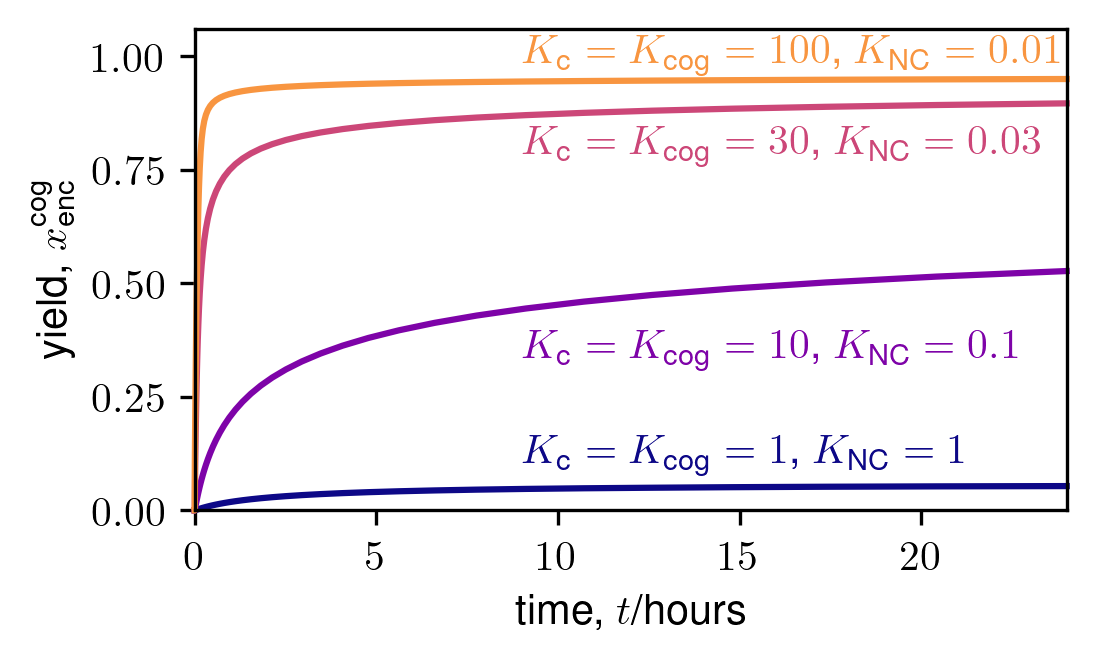}
   \caption{ \textbf{LLPS increases the yield and rate of cognate RNA packaging.} 
    Fraction of cognate RNA molecules in complete capsids as a function of time for indicated values of the subunit and RNA partition coefficients. For all parameter sets except $\Kc=\Kcog=\KNC=1$, nearly all assembly occurs within the condensate. Parameters are: $\rhot=3\upmu$M, $\rhort=\rhocogt+\rhonct=1.5\upmu$M, $\fcog=0.1$, 
    and $\Vr=10^{-3}$. Here $\rhot=N\rhocogt$, so there are just enough subunits to encapsidate every cognate RNA molecule. 
   }
    \label{supp_fig:yield_vs_time_two_species}
\end{figure}

\begin{figure}%[ht]
    \centering
    \includegraphics[width=0.5\linewidth]{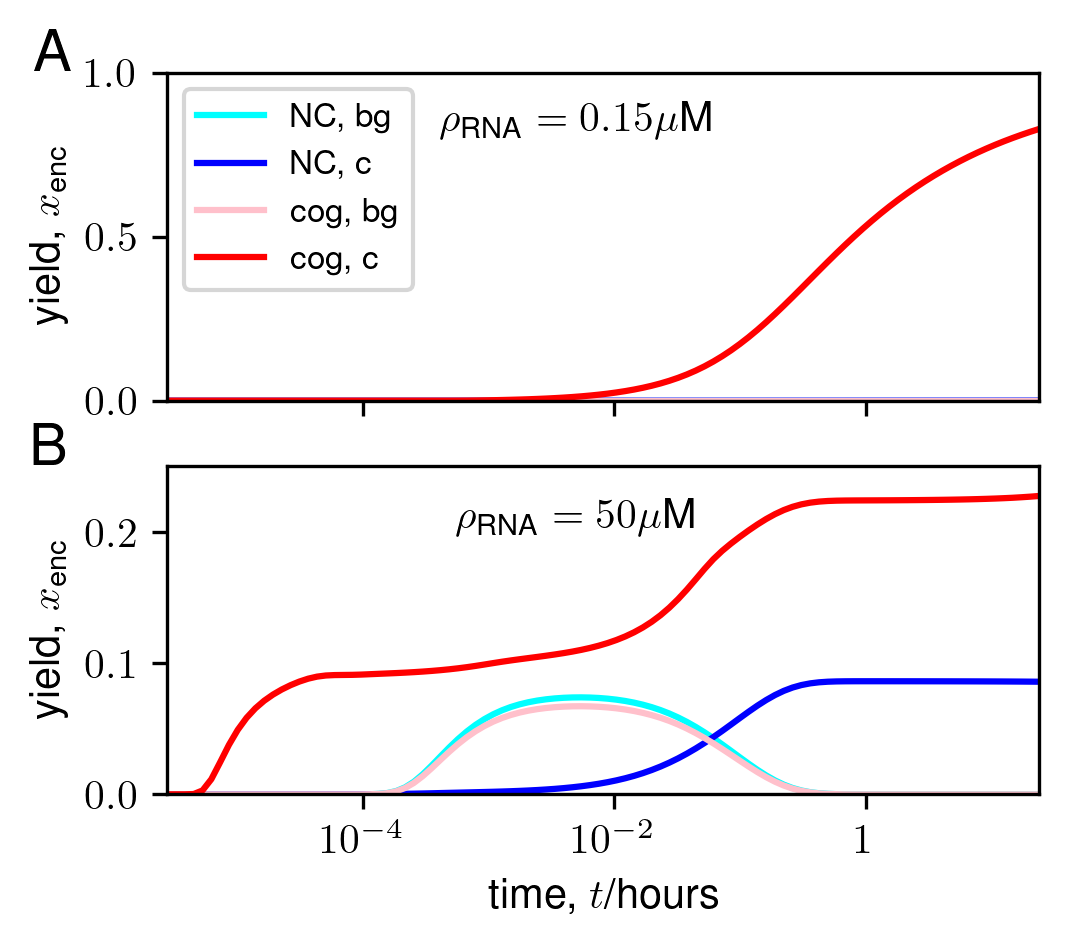}
   \caption{ \textbf{The selectivity and kinetics of RNA packaging depends on total RNA concentration.}
   Fractions of non-cognate (NC) and cognate (cog) RNA in complete capsids in the background (bg) and condensate (c) as a function of time. (A) Moderate total RNA concentration, $\rhort=\rhocogt+\rhonct=0.15\mu$M, corresponding to a typical physiological RNA concentration in mammalian cells (BNID 109916 ~\cite{Milo2010}); (B) High total RNA concentration, $\rhoRNA=5\mu$M.  Note that the $x$-axis is on a logarithmic scale. Cyan, non-cognate RNA in background $\yieldBGnc$; blue, non-cognate RNA in condensate $\yieldCnc$; pink, cognate RNA in background $\yieldBGcog$; red: cognate RNA in condensate $\yieldCcog$. Parameters for panel A are: $\rhot=0.3\upmu$M, $\rhort=0.15\upmu$M, $\Kc=\Kcog=100$, $\KNC=10^{-2}$, $\fcog=0.1$,
   and $\Vr=10^{-3}$. Parameters for panel B are: $\rhot=100\upmu$M, $\Kc=\Kcog=100$, $\KNC=10^{-2}$, $\fcog=0.1$,
   and $\Vr=10^{-3}$. Here $\rhot=N\rhocogt$, so there are just enough subunits to encapsidate every cognate RNA molecule.
   }
    \label{supp_fig:fracs_vs_time_two_species}
\end{figure}

\begin{figure}%[ht]
    \centering
    \includegraphics[width=0.5\linewidth]{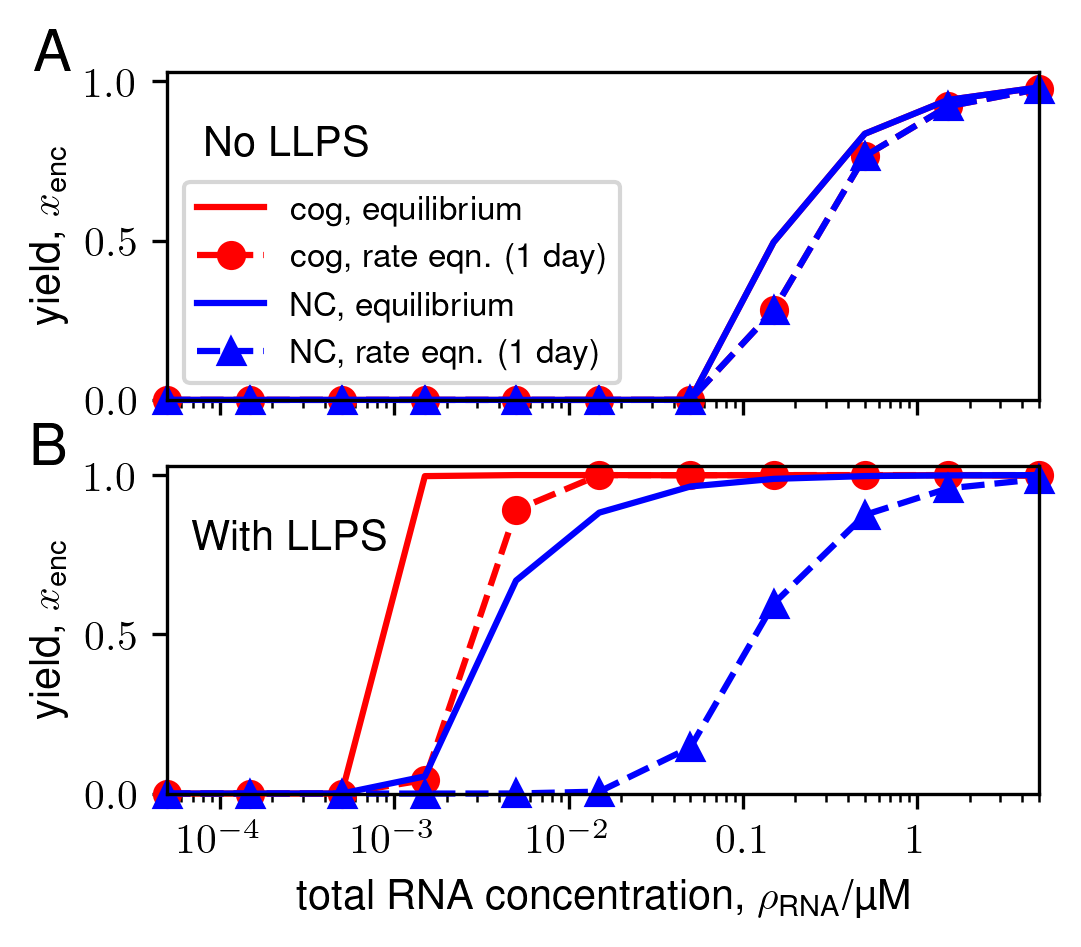}
   \caption{ \textbf{LLPS enables selective packaging at higher stoichiometric ratios $\rhot/N\rhocogt$}.
   (A, B) Fraction of cognate ($\yieldcog$, red, circles) and non-cognate ($\yieldnc$, blue, triangles) RNA in complete capsids versus total subunit concentration $\rhot$, without (A) and with (B) LLPS. Solid lines denote equilibrium results (Section ~\ref{supp_sec:equil_two_species}), while dashed lines with markers correspond to rate equation results (Eq.~\ref{eq:rate_equation}) at $t=1$ day. For these simulations $\fcog=0.1$, $\Vr=10^{-3}$, and $\rhot=N\rhort=10N\rhocogt$, so that there are enough subunits to encapsidate every RNA strand (cognate and non-cognate). (A), $\Kc=\Kcog=\KNC=1$, while for (B), $\Kc=\Kcog=100$ and $\KNC=10^{-2}$. Compared to the results for $\rhot=N\rho_{\text{RNA,cog}}$ (main text Fig.~\ref{fig:yield_two_species}), the range of concentrations where packaging is selective for cognate RNA is smaller.
   }
    \label{supp_fig:yield_two_species_pR=1}
\end{figure}

\begin{figure}%[ht]
    \centering
    \includegraphics[width=0.5\linewidth]{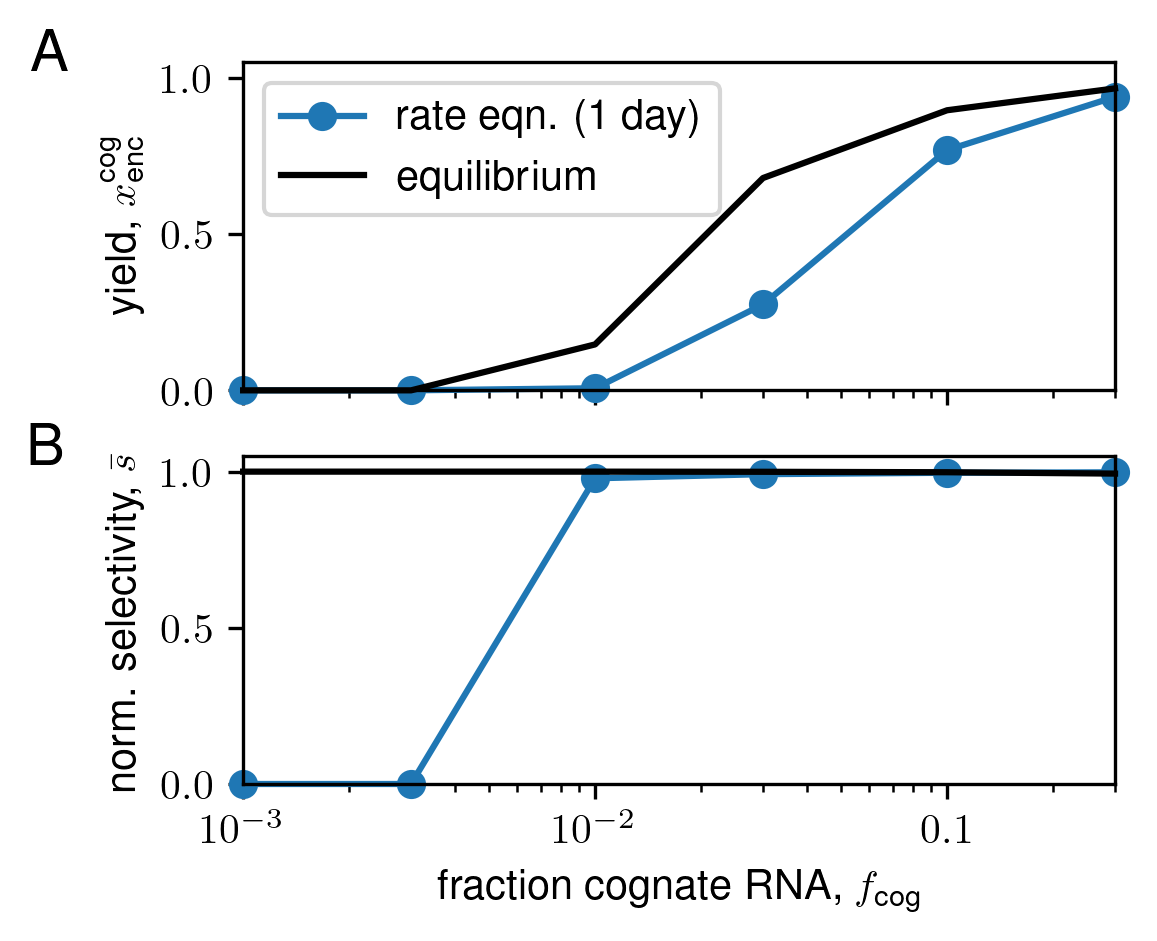}
   \caption{ \textbf{Increasing the fraction of cognate RNA $\fcog$ increases the yield at fixed total RNA concentration. }
   (A) Yield and (B) selectivity as a function of $\fcog$, with  $\Kc=\Kcog=100$, $\KNC=10^{-2}$, $\rhonct=0.1\mu$M, and $\Vr=10^{-3}$. Here $\rhot=N\rhocogt$, so there are just enough subunits to encapsidate every cognate RNA molecule. The black lines are full equilibrium results (Section \ref{supp_sec:equil_two_species}), while the blue lines with circles are rate equation results at $t=1$ day. Note that selectivity remains high even as $\fcog$ is reduced until the yield drops to zero (where, by definition, $\bar{s}=0$).
   }
    \label{supp_fig:yield_selectivity_vary_fcog}
\end{figure}

\begin{figure}%[ht]
    \centering
    \includegraphics[width=0.5\linewidth]{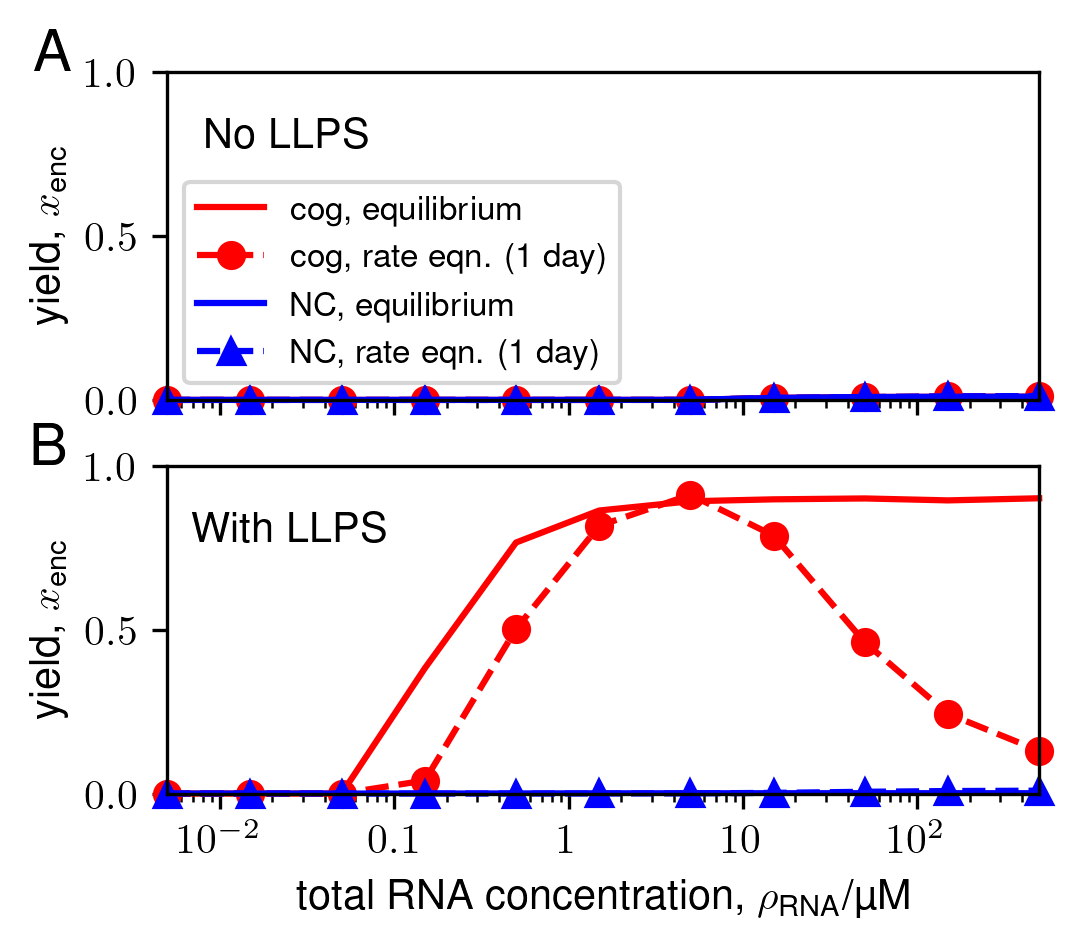}
   \caption{ \textbf{LLPS enables robust selective packaging even at low $\fcog$.}
   (A, B) Cognate ($\yieldcog$ red, circles) and non-cognate ($\yieldnc$, blue, triangles) yields versus total subunit concentration, without (A) and with (B) LLPS. Solid lines denote equilibrium results (Section~\ref{supp_sec:equil_two_species}), while dashed lines with markers correspond to rate equation results (Eq.~\ref{eq:rate_equation}) at $t=1$ day. For these simulations $\fcog=0.01$, $\Vr=10^{-3}$, and $\rhot=N\rho_{\text{RNA,cog}}$, so that there are just enough subunits to encapsidate every cognate RNA strand. (A), $\Kc=\Kcog=\KNC=1$, while for (B), $\Kc=\Kcog=100$ and $\KNC=10^{-2}$. Compared to the results for $\fcog=0.1$ (main text Fig.~\ref{fig:yield_two_species}), the maximum yield is slightly lower and high yields are attained at higher total RNA concentrations.}
    \label{supp_fig:yield_two_species_low_fcog}
\end{figure}

\begin{figure}%[ht]
    \centering
    \includegraphics[width=\linewidth]{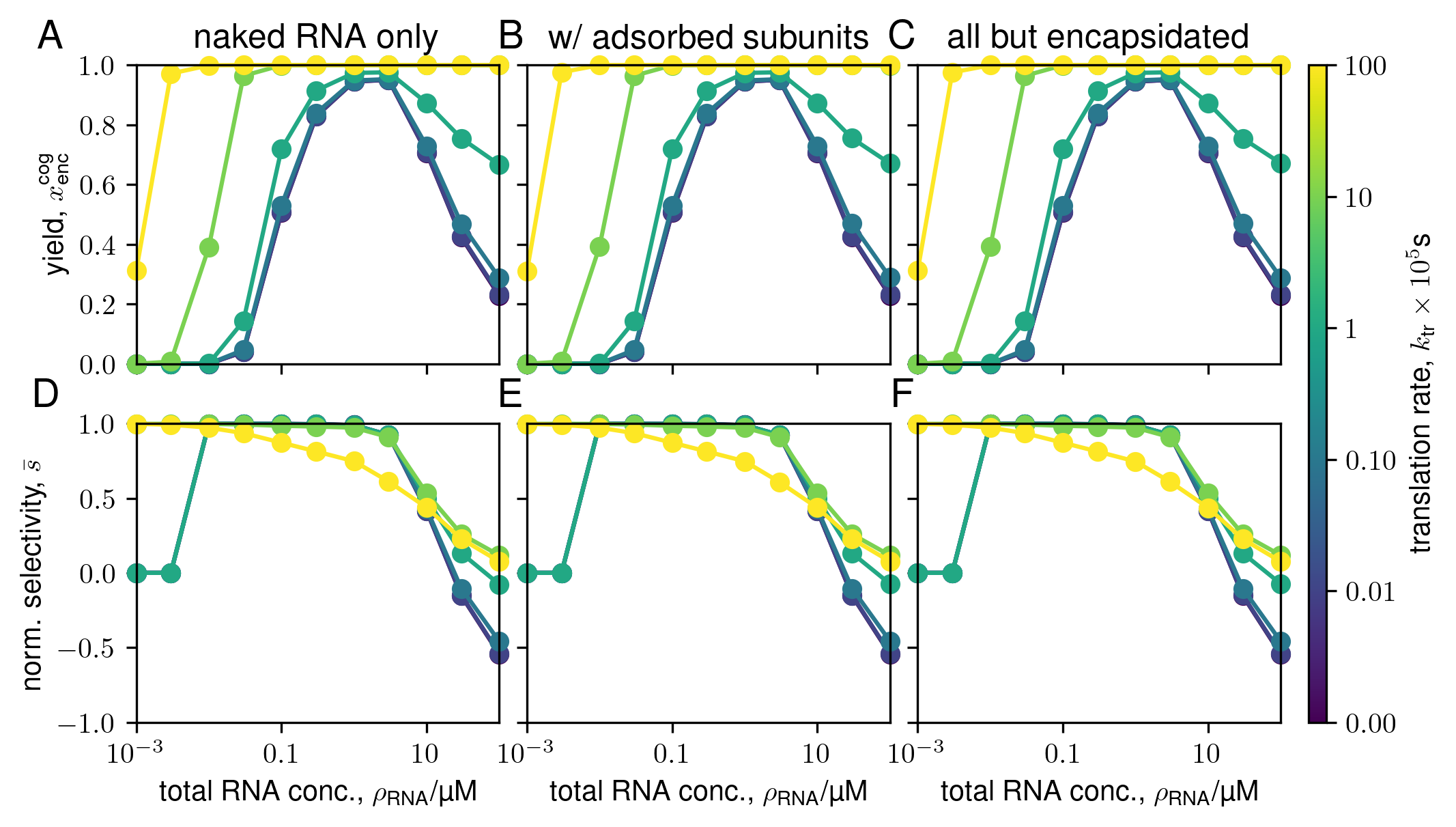}
   \caption{ \textbf{Enhancement of assembly and packaging by subunit translation is robust to changing how accessible RNA is for translation}. The three columns correspond to three different models for translation: (1) in ``naked RNA only'',  only RNA strands with no adsorbed subunits can be translated ($n=m=0$); (2) in ``w/ adsorbed subunits,'' RNA strands with any number adsorbed but no assembled subunits can be translated ($n=0 \ldots N, m=0$); (3) in ``all but encapsidated,'' any RNA strand that is not in a complete capsid can be translated ($m<N$). The top row (panels A-C) shows the yield versus total RNA concentration for different translation rates, and the bottom row (panels D-F) shows the normalized selectivity versus total RNA concentration for different translation rates. The results are nearly identical for the three different methods, showing that the enhancement of assembly by translation is robust to changes in the accessibility of RNA for translation. Parameters are: $\Kc=\Kcog=10^2$, $\KNC=10^{-2}$, $\fcog=0.1$, and $\Vr=10^{-3}$. For $\ktrans=0$, $\rhot=N\rhocogt$, so that there are just enough subunits to encapsidate all cognate RNA strands.
   }
    \label{supp_fig:translation_compare_method}
\end{figure}

\clearpage

\section{Equilibrium Assembly Theory}\label{supp_sec:equil_assem_theory}

In this section we develop theories for the equilibrium yields. Based on previous work \cite{Zlotnick1994,Hagan2014,Hagan2009,Zandi2009,Hagan2021,Zandi2020} we assume that partial capsid intermediates are present at negligible concentrations at equilibrium, thus the only species we consider are free capsid subunits, complete capsids, each of which encloses a single RNA molecule, and RNA molecules with any number of subunits bound up to a maximum number of sites $\np$ ($n=0\ldots \np$). While we set $\np=N$ for the reported results in this work, we allow for any value of $\np$ in this theory for completeness. 

Our derivations rest on two fundamental ideas: conservation of mass, and the law of mass action. Thus, the conceptual framework is simple, although the derivations themselves are somewhat involved. Our basic strategy is to relate yields to the concentrations of unbound subunits (monomers) in the background, via successive applications of mass conservation and mass action. We end up with sets of self-consistent equations for these two quantities, which we solve numerically. We also derive simple, approximate expressions for the yield and selectivity.

For convenience, we define all the variables used in our derivations in Table~\ref{supp_table:equil_vars}.

\begin{table}
\centering
\begin{tabular} { | c | c |}
  \hline
  Parameter &  Description   \\
  \hline
  $\rhot$ & total capsid subunit concentration \\ 
  $\rhocogt$ & total cognate RNA concentration \\ 
  $\rhonct$  &  total non-cognate RNA concentration \\ 
  $\rhorf$ & concentration of ``free'' RNA (not in capsids) \\
  $\rhorfa$ & concentration of ``free'' RNA in phase $\alpha=\{\mbox{c, bg}\}$ \\
  $\rhorna$ & concentration of RNA with $n$ adsorbed subunits in phase $\alpha$ \\
  $\rhoro$ & concentration of ``naked'' RNA (with no adsorbed subunits) \\
  $\rhoroa$ & concentration of naked RNA in phase $\alpha$ \\
  $\rhonuf$ & concentration of ``free'' RNA (not in capsids) of species $\nu=\{\mbox{cog, NC}\}$ \\
  $\rhonufa$ & concentration of ``free'' RNA of species $\nu$ in phase $\alpha=\{\mbox{c, bg}\}$ \\
  $\rhonuna$ & concentration of RNA of species $\nu$ with $n$ adsorbed subunits in phase $\alpha$ \\
  $\rhonuo$ & concentration of ``naked'' RNA (with no adsorbed subunits) of species $\nu$ \\
  $\rhonuoa$ & concentration of naked RNA of species $\nu$ in phase $\alpha$ \\
  $\rhoone$ & concentration of unbound subunits (``monomers'') \\
  $\rhoonea$ & concentration of monomers in phase $\alpha$ \\
  $\rhof$ & concentration of ``free'' subunits (not in complete capsids) \\
  $\rhof^{\alpha}$ & concentration of free subunits in phase $\alpha$ \\
  $\rho_{1\text{RNA}}^{\alpha}$ & concentration of subunits adsorbed to RNA in phase $\alpha$ \\
  $\rho_{1\nu}^{\alpha}$ & concentration of subunits adsorbed to RNA of species $\nu$ in phase $\alpha$ \\
  $\cac$ & critical assembly concentration \\
  $N$ & number of subunits in complete capsid \\
  $\rho_N$ & total capsid concentration \\
  $\rho_{N,\nu}$ & concentration of capsids enclosing RNA of species $\nu$ \\
  $\rho_{N,\nu}^{\alpha}$ & concentration of capsids enclosing RNA of species $\nu$ in phase $\alpha$\\
  $x_N$ & fraction of subunits in complete capsids \\
  $x_N^{\nu}$ & fraction of subunits in complete capsids containing RNA of species $\nu$ \\
  $x_N^{\nu,\alpha}$ & fraction of subunits in complete capsids containing RNA of species $\nu$ in phase $\alpha$ \\
  $\yield$ & yield (fraction of RNA molecules within capsids) \\
  $\rpr$ & stoichiometric ratio of subunits to RNA \\
  $\rprnu$ & stoichiometric ratio of subunits to RNA of species $\nu$ \\
  $\VT$ & total volume \\
  $\Vc$ & condensate volume \\
  $\Vbg$ & background volume \\
  $\Vr$ & condensate size ratio \\
  $\Kc$ & subunit partition coefficient \\
  $\KRNA$ & RNA partition coefficient \\
  $\Kcog$ & cognate RNA partition coefficient \\
  $\KNC$ & non-cognate RNA partition coefficient \\
  $\Gn$ & capsid free energy (not including adsorption free energy) \\
  $\gads$ & adsorption free energy \\
  $\langle \nads \rangle_{\alpha}$ & average number of subunits adsorbed to RNA in phase $\alpha$ \\
  $s$ & selectivity \\
  \hline
\end{tabular}
\caption{Descriptions of variables used in equilibrium theories.}
\label{supp_table:equil_vars}
\end{table}

\subsection{One species of RNA}

Let us first consider the case where there is a single species of RNA. We define the fraction of subunits in complete capsids, $x_N$, as:
\begin{equation}
    x_N = \frac{N\rhon}{\rhot},
\end{equation}
which is proportional to the yield, $\yield$:
\begin{equation}
    \yield = \rpr x_N.\label{eq:yield_to_xn}
\end{equation}
with $\rpr=\rhot/N\rhort$
In what follows, we will first derive an equation for $x_N$; the yield will then be given by Eq.~\ref{eq:yield_to_xn}.

We start by deriving a constraint relating RNA concentrations to the free subunit concentration. Using conservation of mass, we write the concentration of free RNA, $\rhorf$, (i.e. the concentration of RNA molecules not in capsids) as:
\begin{align*}
\rhorf &= \rhort - \rhon \\
       &= \frac{\rhot}{N}\frac{1}{\rpr} - \frac{\rhot}{N}x_N \\
\implies \Aboxed{\rhorf &= \frac{\rhot}{N}\left(\frac{1}{\rpr}-x_N\right).} \numberthis \label{eq:rhorf_llps}
\end{align*}
Invoking conservation of mass across the condensate and background, we have:
\begin{align*}
    \VT \rhorf &= \Vbg \rhorfbg + \Vc \rhorfc \\ 
    \implies \rhorf &= \frac{1}{1+\Vr}\rhorfbg+\frac{\Vr}{1+\Vr}\rhorfc, \numberthis \label{eq:rhorna_cons_c_bg}
\end{align*}
where $\rhorfc$ and $\rhorfbg$ are free RNA concentrations in the condensate (c) and background (bg), respectively. Now, let us use the law of mass action to write $\rhorfa$, where $\alpha=\{\mbox{c, bg}\}$, in terms of the free subunit concentration $\rho_1$ and the concentration of ``naked'' RNA (with no adsorbed subunits), $\rhoroa$. The law of mass action gives:
\begin{equation}
    \rhorna = \rhoroa (\rho_1)^n e^{-n \beta \gads} {\np \choose n},
\end{equation}
where $\rhorna$ is the concentration of RNA with $n$ adsorbed subunits in phase $\alpha$. So, we can write the free RNA concentration in phase $\alpha$ as:
\begin{align*}
\rhorf^{\alpha} &= \sum_{n=0}^{\np} \rhorna \\
                &= \sum_{n=0}^{\np} \rhoroa (\rhoonea)^n e^{-n\beta \gads} {\np \choose n} \\
                &= \rhoroa \left(1+\rhoonea e^{-\beta \gads}\right)^{\np}, \numberthis \label{eq:rhorfa_rhor0a}
\end{align*}
In going from the second to the third line above, we used the binomial theorem. Combining Eqs.~\ref{eq:rhorna_cons_c_bg} and ~\ref{eq:rhorfa_rhor0a} with the definitions $\KRNA=\rhoroc/\rhorobg$ and $\Kc=\rhoonec/\rhoonebg$, we write $\rhorf$ in terms of ``background'' quantities alone:
\begin{equation}
    \rhorf = \frac{1}{1+\Vr}\rhorobg \left(1+\rhoonebg e^{-\beta \gads}\right)^{\np} + \frac{\Vr}{1+\Vr} \KRNA \rhorobg \left(1+\Kc \rhoonebg e^{-\beta \gads}\right)^{\np},
\end{equation}
or rearranging,
\begin{equation}
    \rhorobg = \frac{\rhorf (1+\Vr)}{\left(1+\rhoonebg e^{-\beta \gads}\right)^{\np} + \Vr \KRNA \left(1+\Kc \rhoonebg e^{-\beta \gads}\right)^{\np}}. \label{eq:rhorobg}
\end{equation}
Now, by conservation of RNA mass and RNA partitioning, we can write the total concentration of naked RNA as:
\begin{align*}
\rhoro &= \frac{1}{1+\Vr}\rhorobg + \frac{\Vr}{1+\Vr} \rhoroc \\
       &= \frac{1+\KRNA\Vr}{1+\Vr}\rhorobg \numberthis \label{eq:rhoroToBg}\\
\implies \Aboxed{\rhoro &= \frac{\rhorf (1+\KRNA \Vr)}{\left(1+\rhoonebg e^{-\beta \gads}\right)^{\np}+\Vr\KRNA \left(1+\Kc \rhoonebg e^{-\beta \gads}\right)^{\np}}.} \numberthis \label{eq:rhoro_llps}
\end{align*}
Next, let us focus on the concentration of subunits. We denote the concentration of subunits not in assembled capsids (``free'' subunits, i.e. either monomers or adsorbed to RNA) as $\rhof$, and the concentration of such subunits in phase $\alpha$ as $\rhof^{\alpha}$. By conservation of subunit mass:
\begin{align*}
\rhofbg &= \rhoonebg + \sum_{n=0}^{\np} n\rhornbg \\ 
         &= \rhoonebg + \sum_{n=0}^{\np} n \rhorobg \left(\rhoonebg\right)^n e^{-n\beta \gads} {\np \choose n} \\
         &= \rhoonebg + \rhorobg \left(1+\rhoonebg e^{-\beta \gads}\right)^{\np} \frac{\np \rhoonebg e^{-\beta \gads}}{1+\rhoonebg e^{-\beta \gads}} \\
         &= \rhoonebg + \rhorfbg \langle \nads \rangle_{\text{bg}} \numberthis,
\end{align*}
where:
\begin{equation}
    \langle \nads \rangle_{\alpha} = \frac{\np \rhoonea e^{-\beta \gads}}{1+\rhoonea e^{-\beta \gads}}. \label{eq:nads}
\end{equation} Similarly,
\begin{align*}
    \rhofc &= \rhoonec + \rhorfc \langle \nads \rangle_{\text{c}}. \numberthis
\end{align*}
Now, the total concentration of free subunits is given by:
\begin{align*}
    \VT\rhof &= \Vbg \rhofbg + \Vc \rhofc \\
\implies \rhof &= \frac{1}{1+\Vr}\rhofbg + \frac{\Vr}{1+\Vr}\rhofc \\
               &= \frac{1}{1+\Vr}\left(\rhoonebg + \rhorfbg \langle \nads \rangle_{\text{bg}}\right) + \frac{\Vr}{1+\Vr}\left(\rhoonec + \rhorfc \langle \nads \rangle_{\text{c}}\right) \\
               &= \frac{1+\Kc\Vr}{1+\Vr}\rhoonebg + \frac{1}{1+\Vr} \rhorobg \left(1+\rhoonebg e^{-\beta \gads}\right)^{\np} \langle \nads \rangle_{\text{bg}} + \frac{\Vr}{1+\Vr} \KRNA \rhorobg \left(1+\Kc \rhoonebg e^{-\beta \gads}\right)^{\np} \langle \nads \rangle_{\text{c}}.
\end{align*}
Finally, using Eq. \ref{eq:rhorobg}, we have:
\begin{equation}
    \boxed{\rhof = \rhot(1-x_N) =  \frac{1+\Kc \Vr}{1+\Vr}\rhoonebg + \rho_{1\text{RNA}}^{\text{bg}} + \rho_{1\text{RNA}}^{\text{c}},} \label{eq:rhof_llps}
\end{equation}
where we have defined the concentrations of subunits adsorbed to RNA in the background and condensate as:
\begin{subequations}
\begin{align}
    \rho_{1\text{RNA}}^{\text{bg}} &= \rhorf \langle \nads \rangle_{\text{bg}}\frac{ \left(1+\rhoonebg e^{-\beta \gads}\right)^{\np}}{\left(1+\rhoonebg e^{-\beta \gads}\right)^{\np} + \Vr \KRNA \left(1+\Kc \rhoonebg e^{-\beta \gads}\right)^{\np}} \label{eq:rho1rbg}\\
    \rho_{1\text{RNA}}^{\text{c}} &= \rhorf \langle \nads \rangle_{\text{c}}\frac{ \Vr \KRNA \left(1+\Kc \rhoonebg e^{-\beta \gads}\right)^{\np}}{\left(1+\rhoonebg e^{-\beta \gads}\right)^{\np} + \Vr \KRNA \left(1+\Kc \rhoonebg e^{-\beta \gads}\right)^{\np}}. \label{eq:rho1rc}
\end{align}
\end{subequations}
Eq. \ref{eq:rhof_llps}, combined with Eqs.\ref{eq:rho1rbg}, \ref{eq:rho1rc}, \ref{eq:rhorf_llps}, and \ref{eq:nads}, gives us a single equation with two unknowns, $x_N$ and $\rhoonebg$. To obtain a second equation, we invoke the law of mass action for capsids, combined with Eq. \ref{eq:rhoroToBg}:
\begin{align*}
    \rhonbg &= \rhorobg \left(\rhoonebg\right)^N e^{-N\beta \gads}e^{-\beta G_N} \\
            &= \frac{1+\Vr}{1+\Vr\KRNA }\rhoro \left(\rhoonebg\right)^N e^{-N\beta \gads}e^{-\beta G_N}, \numberthis
\end{align*}
where $\Gn=\sum_{m=0}^N{}^{'}g_m$ is the free energy of an assembled capsid.
Similarly, 
\begin{align*}
    \rhonc &= \rhoroc \left(\rhoonec\right)^N e^{-N\beta \gads}e^{-\beta G_N} \\
    &= \KRNA\Kc^N \frac{1+\Vr}{1+\Vr\KRNA} \rhoro  \left(\rhoonebg\right)^N e^{-N\beta \gads}e^{-\beta G_N} \numberthis.
\end{align*}
So, the total capsid concentration is:
\begin{align*}
    \VT\rhon &= \Vbg \rhonbg + \Vc \rhonc \\
    \implies \rhon &= \frac{1}{1+\Vr} \rhonbg + \frac{\Vr}{1+\Vr}\rhonc \\
    &= \frac{1+\Vr\KRNA\Kc^N}{1+\Vr} \frac{1+\Vr}{1+\Vr\KRNA}\rhoro \left(\rhoonebg\right)^N e^{-N\beta \gads}e^{-\beta G_N} \\
    &= \frac{1+\Vr \KRNA\Kc^N}{1+\Vr\KRNA} \rhoro \left(\rhoonebg\right)^N e^{-N\beta \gads}e^{-\beta G_N}. \numberthis
\end{align*}
The fraction of subunits in complete capsids is therefore:
\begin{equation}
    \boxed{x_N = \frac{N}{\rhot } \frac{1+\Vr \KRNA\Kc^N}{1+\Vr\KRNA} \rhoro \left(\rhoonebg\right)^N e^{-N\beta \gads}e^{-\beta G_N}.} \label{eq:onespecies_yield}
\end{equation}
Eqs. \ref{eq:rhof_llps} and \ref{eq:onespecies_yield}, supplemented with Eqs. \ref{eq:rhorf_llps}, \ref{eq:rhoro_llps}, \ref{eq:rho1rbg}, and \ref{eq:rho1rc}, give us two equations in two unknowns, $x_N$ and $\rhoonebg$. They can be solved numerically; we provide our MATLAB code to do so at [link to be inserted upon publication]. For numerical purposes, it is convenient to take the logarithm of Eq. \ref{eq:onespecies_yield} to avoid computing exponentials of large numbers:
\begin{equation}
    \log{x_N} = \log{N} - \log{\rhot} + \log{\frac{1+\Vr \KRNA\Kc^N}{1+\Vr\KRNA}} + \log{\rhoro} + N\log{\rhoonebg} - N\beta \gads - \beta G_N.
\end{equation}

\subsubsection{Approximate solution and critical assembly concentration ($\cac$)}
\label{sec:ApproximateSolution}

To test the numerical solution and to gain physical insight, we also obtain an approximate for the equilibrium yield, selectivity, and $\cac$.

First, let us write the $\rhoonebg$ in terms of $x_N$ assuming that the concentration of subunits adsorbed to RNA but not assembled is negligible compared to the concentration of monomers and the concentration of subunits in complete capsids:
\begin{align*}
    \rhot &\approx \frac{1}{1+\Vr}\rhoonebg + \frac{\Vr}{1+\Vr}\rhoonec + N \rhon \\
          &= \frac{1+\Vr\Kc}{1+\Vr}\rhoonebg + N\rhon \\
          &= \frac{1+\Vr\Kc}{1+\Vr}\rhoonebg + \rhot x_N \\
    \implies \Aboxed{\rhoonebg &\approx \frac{1+\Vr}{1+\Vr\Kc}\rhot (1-x_N).} \numberthis \label{eq:approx_rho1bg_yield}
\end{align*}
Plugging Eq.~\ref{eq:approx_rho1bg_yield} into Eq. \ref{eq:onespecies_yield} gives:
\begin{align*}
    x_N &= \frac{N}{\rhot} \frac{1+\Vr \KRNA \Kc^N}{1+\Vr \KRNA} \rhoro \left(\frac{1+\Vr}{1+\Vr\Kc}\right)^N \rhot^N (1-x_N)^N e^{-\beta N \gads} e^{-\beta \Gn} \\
    \implies \frac{x_N}{(1-x_N)^N} &= N \rhot^{N-1} \rhoro e^{-N\beta \gads} e^{-\beta \Gn} \frac{1+\Vr \KRNA \Kc^N}{1+\Vr \KRNA}\left(\frac{1+\Vr}{1+\Vr\Kc}\right)^N \\
    \implies \frac{x_N^{1/(N-1)}}{(1-x_N)^{N/(N-1)}} &= \rhot N^{1/(N-1)}\rhoro^{1/(N-1)} e^{-N/(N-1)\beta \gads}e^{-\beta G_N/(N-1)} \\
    &\times \left(\frac{1+\Vr \KRNA \Kc^N}{1+\Vr \KRNA}\right)^{1/(N-1)}\left(\frac{1+\Vr}{1+\Vr\Kc}\right)^{N/(N-1)}. \numberthis
\end{align*}
For $N\gg 1$,
\begin{equation}
    \frac{x_N^{1/N}}{1-x_N} \approx \rhot  e^{-\beta\gads} e^{-\beta \Gn/N} \rhoro^{1/N} \left(\frac{1+\Vr \KRNA \Kc^N}{1+\Vr \KRNA}\right)^{1/N}\frac{1+\Vr}{1+\Vr\Kc},
\end{equation}
assuming excess RNA, $\rhoro\approx \rhort$:
\begin{equation}
    \frac{x_N^{1/N}}{1-x_N} \approx \rhot  e^{-\beta\gads} e^{-\beta \Gn/N} \rhort^{1/N} \left(\frac{1+\Vr \KRNA \Kc^N}{1+\Vr \KRNA}\right)^{1/N}\frac{1+\Vr}{1+\Vr\Kc}.
\end{equation}
The critical assembly concentration is thus:
\begin{equation}
    \boxed{\cac = e^{\beta \gads}e^{\beta \Gn/N} \frac{1+\Kc\Vr}{1+\Vr}\left(\rhort \frac{1+\Vr \KRNA \Kc^N}{1+\Vr\KRNA}\right)^{-1/N}.}
\end{equation}
In the limits where $\rhot$ is either much greater or much less than $\cac$, we obtain simple expressions for $x_N$ and hence the yield (see Eq.~\ref{eq:yield_to_xn}):
\begin{align}  \label{eq:equil_yield_approx_one_species} 
  x_N \approx
    \begin{cases}
        (\cac/\rhot)^N, & \mbox{  for } \rhot\ll \cac  \\
        1-\cac/\rhot, & \mbox{  for } \rhot\gg \cac .
    \end{cases}
\end{align}

\subsection{Two species of RNA} \label{supp_sec:equil_two_species}
Suppose we now have two species of RNA, cognate (cog) and non-cognate (NC), which have partition coefficients $\Kcog$ and $\KNC$. The total concentrations of these different species are denoted $\rhocogt$ and $\rhonct$ respectively. By conservation of mass, these concentrations obey:
\begin{equation}
    \rhonut = \rhonuf + \rhonnu \label{eq:twospecies_rhot}
\end{equation}
where $\nu=\{\mbox{cog, NC\}}$, $\rhonuf$ is the concentration of free RNA (not in capsids) of species $\nu$, and $\rhonnu$ is the concentration of capsids assembled around species-$\nu$ RNA.
Total relative amounts of protein and RNA are set by the ratios:
\begin{equation}
    \rprnu = \frac{\rhot}{N\rhonut}, \;\;\;\nu=\{\mbox{cog, NC}\}.
\end{equation}
The fractions of subunits in complete capsids, $x_N$ (total) and $\xnnu$ (for RNA species $\nu$) are defined as:
\begin{align}
    x_N = \frac{N\rhon}{\rhot} &= \sum_{\nu=\text{NC,cog}}x_N^{\nu}  \label{eq:xn_xnnu}\\
    \xnnu = \frac{N\rhon^{\nu}}{\rhot}.
\end{align}
We also define $x_N^{\nu,\alpha}$ as the fraction of subunits in complete capsids around RNA species $\nu$ in phase $\alpha=\{\mbox{c, bg}\}$:
\begin{subequations}
\begin{align}
    x_N^{\nu,\text{c}} &= \frac{\Vr}{1+\Vr}\frac{N\rhonnuc}{\rhot} \\
    x_N^{\nu,\text{bg}} &= \frac{1}{1+\Vr}\frac{N\rhonnubg}{\rhot}.
\end{align}
\end{subequations}
Following our strategy for the single RNA species case, we first use Eq. \ref{eq:twospecies_rhot} to write:
\begin{align*}
    \rhonuf &= \rhonut - \rhonnu \\
    &= \frac{\rhot}{N\rprnu}-\frac{\rhot}{N}x_N^{\nu} \\
    \implies \Aboxed{\rhonuf &= \frac{\rhot}{N}\left(\frac{1}{\rprnu}-x_N^{\nu}\right).} \numberthis \label{eq:rhonuf_to_xnnu}
\end{align*}
Now, we can write the concentration of free RNA of each species $\nu=\text{NC,cog}$ and phase $\alpha=\text{c,bg}$ as:
\begin{align*}
    \rhonuf^{\alpha} &= \sum_{n=0}^{\np} \rhonun^{\alpha} \\
    &= \sum_{n=0}^{\np}\rhonuo^{\alpha} \left(\rhoone^{\alpha}\right)^n e^{-n\beta \gads} {\np \choose n} \\
    &= \rhonuo^{\alpha}\left(1+\rhoone^{\alpha}e^{-\beta \gads}\right)^{\np}. \numberthis \label{eq:twospecies_rhoafnu}
\end{align*}
Using Eq. \ref{eq:twospecies_rhoafnu} along with RNA mass conservation and the law of mass action, we can write the concentration of free RNA of each species, averaged over background and condensate phases, as:
\begin{align*}
    \VT \rhonuf &= \Vbg \rhonufbg + \Vc \rhonufc \numberthis \\
    \implies \rhonuf &= \frac{1}{1+\Vr}\rhonuobg \left(1+\rhoonebg e^{-\beta \gads}\right)^{\np} + \frac{\Vr}{1+\Vr}\rhonuoc \left(1+\rhoonec e^{-\beta \gads}\right)^{\np} \\
    &= \frac{\rhonuobg}{1+\Vr}\left(\left(1+\rhoonebg e^{-\beta \gads}\right)^{\np} + \Vr \Knu \left(1+\Kc \rhoonebg e^{-\beta \gads}\right)^{\np}\right). \numberthis
\end{align*}
or rearranging, 
\begin{equation}
    \rhonuobg = \frac{\rhonuf (1+\Vr)}{\left(1+\rhoonebg e^{-\beta \gads}\right)^{\np} + \Vr \Knu \left(1+\Kc \rhoonebg e^{-\beta \gads}\right)^{\np}}. \label{eq:rhonu0bg}
\end{equation}
Now, combining Eq.~\ref{eq:rhonu0bg} with conservation of RNA mass and RNA partitioning, the total concentration of naked RNA of each species can be written as:
\begin{align*}
    \rhonuo &= \frac{1}{1+\Vr}\rhonuobg + \frac{\Vr}{1+\Vr}\rhonuoc \\
    &= \frac{1+\Vr \Knu}{1+\Vr}\rhonuobg \numberthis \label{eq:rhonu0_to_rhonu0bg} \\
    \implies \Aboxed{\rhonuo &= \frac{\rhonuf (1+\Vr \Knu)}{\left(1+\rhoonebg e^{-\beta \gads}\right)^{\np} + \Vr \Knu \left(1+\Kc \rhoonebg e^{-\beta \gads}\right)^{\np}}.} \numberthis \label{eq:rhonu0_to_rhonuf}
\end{align*}
Next we focus on subunit concentrations. The concentration of free subunits (not in capsids) in the background, $\rhofbg$, can be written as:
\begin{align*}
    \rhofbg &= \rhoonebg + \sum_{\nu=\text{NC,cog}}\sum_{n=0}^{\np} n \rhonuobg\left(\rhoonebg\right)^n e^{-n\beta \gads} {\np \choose n} \\
            &= \rhoonebg + \sum_{\nu=\text{NC,cog}} \rhonufbg \langle \nads \rangle_{\text{bg}} \\
            &= \rhoonebg + \left(\sum_{\nu=\text{NC,cog}} \rhonuobg \right) \left(1+\rhoonebg e^{-\beta\gads}\right)^{\np} \langle \nads \rangle_{\text{bg}}. \numberthis
\end{align*}
Similarly,
\begin{equation}
\rhofc = \Kc\rhoonebg + \left(\sum_{\nu=\text{NC,cog}} \Knu\rhonuobg \right) \left(1+\Kc\rhoonebg e^{-\beta\gads}\right)^{\np} \langle \nads \rangle_{\text{c}}. 
\end{equation}
Thus, the total concentration of free subunits, $\rhof$, is:
\begin{align*}
\rhof &= \frac{1}{1+\Vr} \rhofbg + \frac{\Vr}{1+\Vr}\rhofc \\
      &= \frac{1+\Vr \Kc}{1+\Vr}\rhoonebg + \frac{\sum_{\nu}\rhonuobg}{1+\Vr}\left(1+\rhoonebg e^{-\beta\gads}\right)^{\np} \langle \nads \rangle_{\text{bg}} + \frac{\Vr \sum_{\nu}\Knu\rhonuobg}{1+\Vr}\left(1+\Kc\rhoonebg e^{-\beta\gads}\right)^{\np} \langle \nads \rangle_{\text{c}}. \numberthis \label{eq:rhof_twospecies}
\end{align*}
Defining:
\begin{subequations}
\begin{align}
    \rho_{1\nu}^{\text{bg}} &= \rhonuf \langle \nads \rangle_{\text{bg}}\frac{ \left(1+\rhoonebg e^{-\beta \gads}\right)^{\np}}{\left(1+\rhoonebg e^{-\beta \gads}\right)^{\np} + \Vr \Knu \left(1+\Kc \rhoonebg e^{-\beta \gads}\right)^{\np}} \label{eq:twospecies_rho1rbg}\\
    \rho_{1\nu}^{\text{c}} &= \rhonuf \langle \nads \rangle_{\text{c}}\frac{ \Vr \Knu \left(1+\Kc \rhoonebg e^{-\beta \gads}\right)^{\np}}{\left(1+\rhoonebg e^{-\beta \gads}\right)^{\np} + \Vr \Knu \left(1+\Kc \rhoonebg e^{-\beta \gads}\right)^{\np}}, \label{eq:twospecies_rho1rc}
\end{align}
\end{subequations}
and combining Eq.~\ref{eq:rhof_twospecies} with Eq.~\ref{eq:rhonu0bg}, we have:
\begin{equation}
    \boxed{\rhof = \rhot(1-x_N) = \frac{1+\Vr\Kc}{1+\Vr}\rhoonebg + \sum_{\nu=\mbox{cog,NC}}\left(\rho_{1\nu}^{\text{bg}} + \rho_{1\nu}^{\text{c}}\right).} \label{eq:twospecies_rhof}
\end{equation}
Finally, let us invoke mass action for capsids:
\begin{equation}
    \rhonnubg = \rhonuobg\left(\rhoonebg\right)^N e^{-N\beta \gads} e^{-\beta G_N}.
\end{equation}
The total concentration of capsids is:
\begin{align*}
    \rhon &= \frac{1}{1+\Vr}\sum_{\nu}\rhonnubg + \frac{\Vr}{1+\Vr}\sum_{\nu}\rhonnuc \\
          &= \frac{1}{1+\Vr}\left(\rhoonebg\right)^N e^{-N\beta \gads} e^{-\beta G_N} \left(\sum_{\nu}\rhonuobg\right) + \frac{\Vr}{1+\Vr}\Kc^N\left(\rhoonebg\right)^N e^{-N\beta \gads} e^{-\beta G_N} \left(\sum_{\nu}\Knu\rhonuobg\right) \\
          &= \frac{1}{1+\Vr}\left(\rhoonebg\right)^N e^{-N\beta \gads} e^{-\beta G_N}\left(\sum_{\nu}\rhonuobg \left(1+\Vr \Knu \Kc^N\right)\right) \\
          &= \sum_{\nu}\left(\frac{1+\Vr \Knu\Kc^N}{1+\Vr \Knu}\rhonuo\right)\left(\rhoonebg\right)^N e^{-N\beta \gads} e^{-\beta G_N}, \numberthis
\end{align*}
where we invoked Eq.~\ref{eq:rhonu0_to_rhonu0bg} in going from the third to the fourth line.
The fraction of subunits in complete capsids, $x_N$, is therefore:
\begin{equation}
    \boxed{x_N = \frac{N}{\rhot}\sum_{\nu=\mbox{cog,NC}}\left(\frac{1+\Vr \Knu\Kc^N}{1+\Vr \Knu}\rhonuo\right)\left(\rhoonebg\right)^N e^{-N\beta \gads} e^{-\beta G_N}.} \label{eq:twospecies_yield}
\end{equation}
Eqs. \ref{eq:twospecies_rhof} and \ref{eq:twospecies_yield}, supplemented with Eqs.~\ref{eq:rhonu0_to_rhonuf},~\ref{eq:rhonuf_to_xnnu},~\ref{eq:xn_xnnu},~\ref{eq:twospecies_rho1rbg}, and~\ref{eq:twospecies_rho1rc},  constitute three equations in three unknowns, $x_N$, $x_N^{\text{cog}}$, and $\rhoonebg$, and can be solved numerically (see our MATLAB code at [link to be inserted upon publication]). For numerical stability it is more convenient to take the logarithm of Eq. \ref{eq:twospecies_yield}:
\begin{equation}
    \log{x_N} = \log{N} - \log{\rhot} + \log{\left(\sum_{\nu=\mbox{cog,NC}}\frac{1+\Vr \Knu\Kc^N}{1+\Vr \Knu}\right)} + \log{\left(\sum_{\nu=\mbox{cog,NC}}\rhonuo\right)} + N\log{\rhoonebg} - N\beta \gads - \beta G_N.
\end{equation}
In practice, we cast the problem of solving this set of nonlinear equations for three unknowns as a nonlinear least-squares problem. We write the equations in the form $\mathbf{h}(x_N,x_N^{\text{cog}},\rhoonebg)=\mathbf{0}$, where $\mathbf{h}$ is a three-dimensional vector. We then use MATLAB's nonlinear least-squares solver to minimize the components of $\mathbf{h}$. The advantage of this approach is that it allows us to impose constraints, such that concentrations are forced to be $>0$ and fractions of subunits in capsids are forced to be between 0 and 1; we found that other, simpler approaches to solving the equations failed to satisfy these constraints. In addition, we found that the ability of the solver to find a good least-squares fit was sensitive to the initial guess for the parameters $x_N,x_N^{\text{cog}},\rhoonebg$. We therefore start the solver from a range of initial guesses for the parameters, compute the sum of squared residuals, and pick the parameters for which this sum is minimum.

Once $x_N,x_N^{\text{cog}}$, and $\rhoonebg$ have been obtained, $x_N^{\text{NC}}$ can be computed via Eq.~\ref{eq:xn_xnnu}. Furthermore, we can compute $x_N^{\nu,\alpha}$, $\nu=\{\mbox{cog, NC}\}$, $\alpha=\{\mbox{c, bg}\}$ (Eq.~\ref{eq:yield_cog_nc_c_bg}) by using conservation of mass between phases:
\begin{equation}
    x_N^{\nu} = x_N^{\nu,\text{c}} + x_N^{\nu,\text{bg}}
\end{equation}
combined with the capsid partitioning equation:
\begin{equation}
    x_N^{\nu,\text{c}}/x_N^{\nu,\text{bg}} = \Vr\Kc^N\Knu,
\end{equation}
which follows from Eq.~\ref{eq:partitionCoefIntermediate}. Thus, e.g., 
\begin{equation}
    x_N^{\nu, \text{c}} = \frac{x_N^{\nu}}{1+(\Vr\Kc^N\Knu)^{-1}}.
\end{equation}

\subsubsection{Approximate yield}
As in the single-species case, we write the background concentration in terms of the $x_N$:
\begin{equation}
    \rhoonebg \approx \frac{1+\Vr}{1+\Vr\Kc}\rhot (1-x_N).
\end{equation}

Plugging this into Eq. \ref{eq:twospecies_yield} gives:
\begin{align*}
    x_N &= \frac{N}{\rhot}\sum_{\nu}\left(\frac{1+\Vr \Knu\Kc^N}{1+\Vr \Knu}\rhonuo\right) \left(\frac{1+\Vr}{1+\Vr\Kc}\right)^N \rhot^N (1-x_N)^N e^{-\beta N \gads} e^{-\beta \Gn} \\
    \implies \frac{x_N}{(1-x_N)^N} &= N \rhot^{N-1} e^{-N\beta \gads} e^{-\beta \Gn} \left(\frac{1+\Vr}{1+\Vr\Kc}\right)^N \sum_{\nu}\left(\frac{1+\Vr \Knu \Kc^N}{1+\Vr \Knu}\rhonuo\right) \\
    \implies \frac{x_N^{1/(N-1)}}{(1-x_N)^{N/(N-1)}} &= \rhot N^{\frac{1}{N-1}}e^{-\frac{N}{N-1}\beta \gads}e^{-\frac{\beta G_N}{N-1}} \left(\frac{1+\Vr}{1+\Vr\Kc}\right)^{\frac{N}{N-1}}\left(\sum_{\nu}\frac{1+\Vr \Knu \Kc^N}{1+\Vr \Knu}\rhonuo\right)^{\frac{1}{N-1}}. \numberthis
\end{align*}
For $N\gg 1$,
\begin{equation}
    \frac{x_N^{1/N}}{1-x_N} \approx \rhot  e^{-\beta\gads} e^{-\beta \Gn/N} \frac{1+\Vr}{1+\Vr\Kc} \left(\sum_{\nu}\frac{1+\Vr \Knu \Kc^N}{1+\Vr \Knu}\rhonuo\right)^{\frac{1}{N}},
\end{equation}
assuming excess RNA, $\rhonuo\approx \rhonut$:
\begin{equation}
    \frac{x_N^{1/N}}{1-x_N} \approx \rhot  e^{-\beta\gads} e^{-\beta \Gn/N} \frac{1+\Vr}{1+\Vr\Kc} \left(\sum_{\nu}\frac{1+\Vr \Knu \Kc^N}{1+\Vr \Knu}\rhonut\right)^{\frac{1}{N}},
\end{equation}
The critical assembly concentration is thus:
\begin{equation}
    \boxed{\cac = e^{\beta \gads}e^{\beta \Gn/N} \frac{1+\Vr\Kc}{1+\Vr}\left(\sum_{\nu=\mbox{cog,NC}}\frac{1+\Vr \Knu \Kc^N}{1+\Vr \Knu}\rhonut\right)^{-\frac{1}{N}},}
\end{equation}
and $x_N$ (and hence the yield, via Eq.~\ref{eq:yield_to_xn},) is given by:
\begin{align}  \label{eq:equil_yield_approx_one_species} 
  x_N \approx
    \begin{cases}
        (\cac/\rhot)^N, & \mbox{  for } \rhot\ll \cac  \\
        1-\cac/\rhot, & \mbox{  for } \rhot\gg \cac .
    \end{cases}
\end{align}

\subsubsection{Selectivity}
We define selectivity, $s$, as the ratio of cognate to non-cognate capsid concentrations:
\begin{equation}
    s = \frac{\rho_{N,\text{cog}}}{\rho_{N,\text{NC}}}.
\end{equation}
We will derive an approximate expression for $s$, assuming that the concentration of adsorbed but unassembled subunits on RNA is negligible. First, recall that the concentrations of capsid around either species $\nu=\{\mbox{cog,NC}\}$ of RNA depend on their values in both the background and compartment:
\begin{equation}
    \rho_{N,\nu} = \frac{1}{1+\Vr}\rhonnubg + \frac{\Vr}{1+\Vr}\rhonnuc.
\end{equation}
We thus need approximate expressions for the capsid concentration of either species in either phase. To do so we appeal to the law of mass action:
\begin{equation}
    \rho_{N,\nu}^{\alpha} = \rhonuf^{\alpha}\left(\rhoone^{\alpha}\right)^N e^{-\beta (\Gn+N\gads)}.
\end{equation}
Let us first get an approximate expression for $\rhoone^{\alpha}$. Since we assume that the concentration of adsorbed but unassembled subunits is negligible, we can write:
\begin{align*}
    \rhot &= \frac{1}{1+\Vr}\rhoonebg + \frac{\Vr}{1+\Vr}\rhoonec + N\rhon \\
    &= \frac{1+\Vr\Kc}{1+\Vr}\rhoonebg + \rhot x_N \\
    \implies \rhoonebg &= \frac{1+\Vr}{1+\Vr\Kc}\rhot (1-x_N) \text{     or    } \rhoonec = \Kc \frac{1+\Vr}{1+\Vr\Kc}\rhot (1-x_N).
\end{align*} 
Defining:
\begin{equation}
    \psi = \frac{1+\Vr}{1+\Vr\Kc},
\end{equation}
we have:
\begin{subequations}
    \begin{align}
        \rhoonebg &= \psi \rhot (1-x_N) \\
        \rhoonec &= \Kc \psi \rhot (1-x_N).
    \end{align}
\end{subequations}
Next, we need an approximate expression for $\rho_{N,\nu}^{\alpha}$. To do so, note that since we assume an excess of RNA, the concentration of capsid-bound RNA will be neglibile compared to that of free RNA. Defining:
\begin{equation}
    \psi_{\nu} = \frac{1+\Vr}{1+\Vr \Knu},
\end{equation}
we have:
\begin{align}
    \rhonut &= \frac{1}{1+\Vr}\rhonufbg + \frac{\Vr}{1+\Vr}\rhonufc \\
           &= \frac{1+\Vr \Knu}{1+\Vr}\rhonufbg \\
           &= \frac{1}{\Knu}\frac{1+\Vr \Knu}{1+\Vr}\rhonufc \\
           &= \Knu^{-1}\psi_{\nu}^{-1} \rhonufc.
\end{align}
Therefore, we can write mass action for each species and phase as:
\begin{subequations}
    \begin{align}
        \rho_{N,\text{cog}}^{\text{c}} &= \Kcog \psi_{\text{cog}} \Kc^N \psi^N \rhot^N \rhocogt (1-x_N)^N e^{-\beta (\Gn+N\gads)} \\
        \rho_{N,\text{cog}}^{\text{bg}} &= \psi_{\text{cog}} \psi^N \rhot^N \rhocogt (1-x_N)^N e^{-\beta (\Gn+N\gads)} \\
        \rho_{N,\text{NC}}^{\text{c}} &= \KNC \psi_{\text{NC}} \Kc^N \psi^N \rhot^N (1-x_N)^N e^{-\beta (\Gn+N\gads)} \\
        \rho_{N,\text{NC}}^{\text{bg}} &= \psi_{\text{NC}} \psi^N \rhot^N (1-x_N)^N e^{-\beta (\Gn+N\gads)}.
    \end{align}
\end{subequations}
So, the selectivity is given by:
\begin{align*}
    s &= \frac{\rho_{N,\text{cog}}^{\text{bg}} + \Vr \rho_{N,\text{cog}}^{\text{c}}}{\rho_{N,\text{NC}}^{\text{bg}} + \Vr \rho_{N,\text{NC}}^{\text{c}}} \\
    &= \frac{\psi_{\text{cog}}\rho_{\text{cog},T} +  \Vr\Kcog \Kc^N \psi_{\text{cog}}\rho_{\text{cog},T}}{\psi_{\text{NC}}\rho_{\text{NC},T} +  \Vr\KNC \Kc^N \psi_{\text{NC}}\rho_{\text{NC},T}} \\
    &= \boxed{\frac{\rho_{\text{cog},T} (1+\Vr \KNC)(1+\Vr \Kcog \Kc^N)}{\rho_{\text{NC},T}(1+\Vr \Kcog)(1+\Vr \KNC \Kc^N)}.} \numberthis
\end{align*}
This expression simplifies further in the limit where all assembly in the compartment is around cognate RNA, and all assembly in the background is around non-cognate RNA. In this limit, $\rho_{N,\text{NC}}^{\text{c}}\approx 0$ and $\rho_{N,\text{cog}}^{\text{bg}}\approx 0$, so:
\begin{equation}
    \boxed{s\approx \frac{\rho_{\text{cog},T} (1+\Vr \KNC)(\Vr \Kcog \Kc^N)}{\rho_{\text{NC},T}(1+\Vr \Kcog)}.}
\end{equation}

\section{Timescale for nucleation in the background}\label{supp_sec:timescales}

At high subunit and RNA concentrations, capsid nucleation on non-cognate RNA may occur in the background before subunits have time to diffuse into the condensate. This reduces selectivity, and in the case of excess non-cognate RNA ($N \rhot / (1-\fcog) \rhoRNA \ll 1$)  subsequent growth of nuclei on non-cognate RNA will deplete subunits resulting in the monomer starvation kinetic trap \cite{Hagan2014}.  
In this section we estimate the threshold subunit concentration above which this trap occurs as a function of the other control parameters, by computing the concentration at which the timescale for subunits to diffuse into the condensate, $\tauD$, is equal to the timescale for nucleation on non-cognate RNA.

 We assume that, in the background, subunit adsorption and elongation times are negligible compared to the nucleation timescale around non-cognate RNA, $\taunucnc$. We also assume that assembly in the condensate is fast compared to nucleation in the background. 

We seek the total subunit concentration $\rhot$ for which:
\begin{equation}
    \taunucnc < \tauD.
\end{equation}
We previously derived an approximate expression for $\tauD$~\cite{Frechette2025}:
\begin{equation}
    \tauD \approx \frac{\Vc(\Vr+1/\Kc)+x_N\VT}{4\pi D_1 R_{\text{c}}},
\end{equation}
where $R_{\text{c}}$ is the radius of a spherical condensate of volume $\Vc$. To estimate $\taunucnc$, we compute the concentration of adsorbed subunits, $\rhoads$, as:
\begin{equation}
    \rhoads = \frac{\rhot e^{-\beta \gads}}{1+\rhot e^{-\beta \gads}}.
\end{equation}
The nucleation time is then given by:~\cite{Hagan2008, Hagan2010}
\begin{equation}
    \taunucnc^{-1} = \fElong \rhoads^{\nnuc} e^{-\beta (\nnuc-1)\gNuc}
\end{equation}
Using the values given in Table~\ref{table:parameters} and assuming high yield ($x_N\approx 1$), we find that $\taunucnc < \tauD$ when $\rhot\gtrsim 20\upmu M$. This value predicts reasonably well the value of $\rhot$ in Fig.~\ref{fig:yield_two_species}B above which cognate yield drops.

\section{Estimate of $\fElong$} \label{supp_sec:fElong_estimate}

While Dykeman et al.~\cite{Dykeman2013} used a range $\fElong = \{100,10^6\}\text{s}^{-1}$, we can make a crude estimate of $\fElong$ as follows. Assuming that a subunit binding site on the RNA is on the order of subunit size, $\lSite=5$nm or about 10 nucleotides, we can estimate the local concentration $\rhoSite$ given that a subunit is bound at the adjacent site as one subunit within a spherical volume with diameter $\lSite$, which gives $\rhoSite \approx 2.5\times10^{-2}$M. We recently estimated \cite{Mohajerani2022}  the association constant for HBV subunits from the lag times measured with \textit{in vitro} assembly experiments \cite{Selzer2014} as $\fSol \approx 10^7 / \text{M} \cdot \text{s}$. However, given that diffusion is 10-fold slower within a cell than in a typical aqueous solution, we use $\fSol = 10^6/ \text{M} \cdot \text{s}$. Finally, this gives $\fElong = \fSol \rhoSite \approx 10^4/$s, which is the value we use throughout this work. Interestingly, it is in the middle of the range that Dykeman et al.~\cite{Dykeman2013} considered.

\bibliography{RNA_LLPS_references}